\documentclass[preprint2]{aastex}

\usepackage{subfig,subtab}

\def\etal{{et\,al.}}
\def\farcs{\hbox{$.\!\!^{\prime\prime}$}}  
\def\degs{\ifmmode ^{\circ}\else$^{\circ}$\fi}
\def\amin{\ifmmode ^{\prime}\else$^{\prime}$\fi}
\def\asec{\ifmmode ^{\prime\prime}\else$^{\prime\prime}$\fi}
\newbox\grsign \setbox\grsign=\hbox{$>$}
\newdimen\grdimen \grdimen=\ht\grsign
\newbox\laxbox \newbox\gaxbox
\setbox\gaxbox=\hbox{\raise.5ex\hbox{$>$}\llap
     {\lower.5ex\hbox{$\sim$}}}\ht1=\grdimen\dp1=0pt
\setbox\laxbox=\hbox{\raise.5ex\hbox{$<$}\llap
     {\lower.5ex\hbox{$\sim$}}}\ht2=\grdimen\dp2=0pt
\def\gax{$\mathrel{\copy\gaxbox}$}
\def\lax{$\mathrel{\copy\laxbox}$}

\slugcomment{Version of \today}

\shorttitle{GROND}
\shortauthors{Greiner et al.}

\begin{document}

\title{GROND - a 7-channel imager}


\author{J. Greiner,
W. Bornemann,
C. Clemens,
M. Deuter,
G. Hasinger,
M. Honsberg,
H. Huber,
S. Huber,
M. Krauss\altaffilmark{1}
T. Kr\"uhler,
A. K\"{u}pc\"{u} Yolda\c{s},
H. Mayer-Hasselwander,
B. Mican,
N. Primak,
F. Schrey,
I. Steiner\altaffilmark{2},
G. Szokoly\altaffilmark{3},
C.C. Th\"one\altaffilmark{4},
A. Yolda\c{s}
}
\affil{Max-Planck-Institut f\"ur extraterrestrische Physik, 85740 Garching,
    Germany}
\email{jcg, wab, cclemens, grh, hhuber, shuber, kruehler, ayoldas, H-M-H, 
b.mican, prima, fzs, yoldas@mpe.mpg.de}

\and

\author{S. Klose, U. Laux, J. Winkler}
\affil{Th\"uringer Landessternwarte, Sternwarte 5, 07778 Tautenburg, Germany}
\email{laux, klose, john@tls-tautenburg.de}

\altaffiltext{1}{present address: ATV Technologie GmbH, D-85591 Vaterstetten,
Johann-Sebastian-Bach-Str. 38; email: markus.krauss@atv-tech.de}
\altaffiltext{2}{present address: Ruhr-Universit\"at Bochum,
Astronomisches Institut, Universit\"atsstr. 150, 44780 Bochum;  email:
    isteiner@astro.ruhr-uni-bochum.de}
\altaffiltext{3}{present address: Institute of Physics, E\"otv\"os University,
    P\'azm\'any P. s. 1/A, 1117 Budapest, Hungary; szgyula@elte.hu}
\altaffiltext{4}{present address: Dark Cosmology Center, Niels Bohr Institute,
   Univ. of Copenhagen, Juliane Maries Vej 30, DK-2100 Kobenhavn; 
   cthoene@dark-cosmology.dk}

\begin{abstract}
We describe the construction of GROND, a 7-channel imager,
primarily designed for rapid observations of gamma-ray burst afterglows. 
It allows simultaneous imaging in the Sloan $g'r'i'z'$ and near-infrared
$JHK$ bands. GROND was commissioned at the MPI/ESO 2.2\,m 
telescope at La Silla (Chile) in April 2007, and first results of
its performance and calibration are presented.
\end{abstract}


\keywords{instrumentation: detectors,  techniques: photometric}

\section{Introduction}

Simultaneous imaging in different filter-bands is of interest
in a variety of astrophysical areas. The primary aim is to
measure the spectral energy distribution or its evolution in 
variable objects in order
to uncover the underlying emission mechanism.
Examples are, among others,
(1) monitoring of all kinds of variable stars (flare stars, cataclysmic
  variables,
  X-ray binaries) to determine the outburst mechanisms and
  differentiate between  physical state changes and changes induced
  by geometrical variations, like eclipses;
(2) monitoring of AGN to understand the physical origin of the
  observed variability;
(3) determining the inclination of X-ray heated binaries \cite{orosz} 
(4) mapping of galaxies to study the stellar population;
(5) multi-color light curves of supernovae \cite{tpm06}; 
(6) differentiate achromatic microlensing events \cite{pac86}
   from other variables with similar light curves;
(7) identifying objects with peculiar spectral energy distributions, e.g. 
  photometric redshift surveys for high-$z$ active galactic nuclei, or
  identifying brown dwarfs;
(8) follow-up observations of transiting extrasolar planets \cite{jha00}; or
(9) mapping of reflectance of solar system bodies as a function of 
their rotation to map their surface chemical composition \cite{jew02}.

A new need for multi-band imaging
arose with the observation of a large number of gamma-ray burst
(GRB) afterglows with the {\it Swift} satellite \cite{geh04}. 
With its much more sensitive
instruments it detects GRBs over a very wide redshift range. Since
intermediate to high-resolution spectroscopy to measure the physical 
conditions of the burst environment \citep[e.g.][]{vre07}
is constrained to the first few hours after
a GRB explosion, a rapid determination of the redshift became important.
This is best done with multi-band photometry (until integral field units
have grown to several arcmin field-of-views) and deriving a photometric
redshift based on the Ly$\alpha$ break \cite{lr00}.

Previous and current instruments with simultaneous imaging capability
in different filter bands include 
ANDICAM (A Novel Double-Imaging CAMera; two channels, one for visual, 
the other for near-infrared \cite{dep98}, presently operated at a 1.3\,m
telescope),
BUSCA (Bonn University Simultaneous CAmera; four visual channels \cite{rei99},
presently operated at the 2.2\,m telescope at Calar Alto),
HIPO (High-speed Imaging optical Photometer for Occultations; two visual 
channels \cite{dun04},
to be operated on SOFIA, the Stratospheric Observatory For Infrared Astronomy),
MITSuME (Multicolor Imaging Telescopes for Survey and Monstrous Explosions;
three channels with fixed bands g\amin, R$_{\rm C}$ and I$_{\rm C}$
\cite{kky07}, operated at a 50 cm telescope),
TRISPEC (Triple Range Imager and Spectrograph; three channels with one CCD and 
two near-IR detectors and wheels for filters, grisms and Wollaston prisms
\citep{wny05}), 
SQIID (Simultaneous Quad Infrared Imaging Device; $JHK$ and narrow-band $L$
filters in front of  individual 512X512 quadrants of an ALADDIN InSb array,
designed for the f/15 Cassegrain foci of the KPNO 2.1-m and 4-m telescopes
\citep{edf92}),
ULTRACAM (ULTRAfast, triple-beam CCD CAMera for high-speed astrophysics 
\cite{dhi07}; portable instrument which has been used, among others, at
the Very Large Telescopes at ESO, or the William Herschel Telescope,
Canary Islands).

Here we describe the design ($\S$2) and performance ($\S$6) 
of a 7-channel imager,
called GROND ({\bf G}amma-{\bf R}ay Burst {\bf O}ptical and 
{\bf N}ear-Infrared {\bf D}etector),
which was specifically designed for GRB afterglow observations.
We also mention some basics of the operation scheme ($\S$3),
related software ($\S$4), and
the changes to the telescope infrastructure which
were implemented to use GROND for rapid follow-up observations ($\S$5).

\section{Instrument Design}

\subsection{Scientific requirements}

The primary goal of identifying GRB afterglows and measuring their redshift 
led to the concept of a camera which allows observations in several 
filters throughout the optical and near-infrared region at the same time.
The simultaneity is dictated by the fact that a typical GRB afterglow
fades by about 2--3 mag between 5--10 min after the GRB, and by another 4 mag
in the following 50 min (e.g. Kann \etal\ 2008). 
Rapidly determining the photometric redshift of GRBs at $z>5$ requires
multi-band photometry in at least 3--4 bands. In order to have
a stable lever arm to determine (i) the intrinsic power law slope of the
continuum emission and (ii) the galactic foreground as well as GRB host
intrinsic extinction, near-infrared bands up to $K$ are essential.
Extending the wavelength coverage beyond $K$ by including, e.g. an $L$ channel,
was seriously considered during the design phase, but both the prospects
of detecting a GRB in the $L$ band as well as the substantial, additional
technical constraints due to the required lower temperatures led us to
drop this possibility.
Going as blue as possible is warranted by detecting the Ly-break down to
redshifts of $\approx$3; however, also including a $U$ or $u'$ band turned
out to be difficult due to space and logistics constraints. Leaving out
one or two bands within our wavelength band would increase the error in
the photo-$z$ determination by a factor of three at least (for the redshift
range covered by that band).
Thus, we decided to
use four bands in the visual, plus the standard $JHK$ bands in the 
near-infrared (NIR).

The field-of view (FOV) of the camera should be large enough to cover
the typical error boxes of gamma-ray bursts, but on the other hand
have a pixel scale less than the mean seeing  to allow
accurate photometry. Given the typical brightness of GRB afterglows,
a reasonably large telescope with permanent access
is another important requirement. Finally, as GROND would act as
pathfinder to pre-select ``interesting'' GRBs for more detailed
follow-up observations, a nearby 8--10\,m telescope
was considered advantageous in the search for the final host telescope.

\begin{figure}[ht]
\hspace{-0.4cm}\includegraphics[width=1.05\columnwidth]{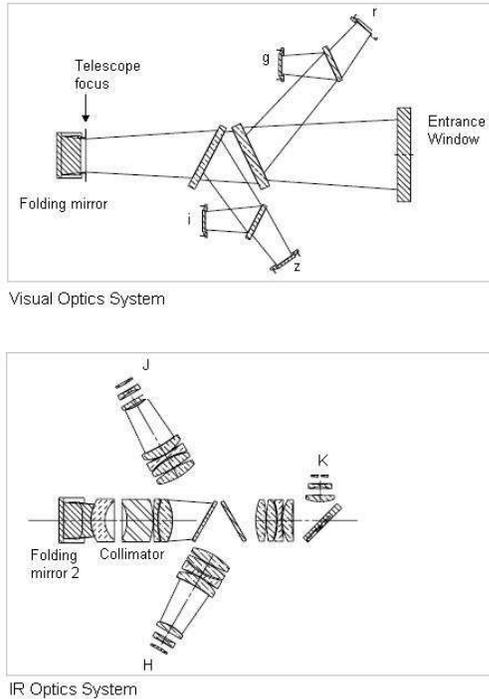}
\caption{Layout of the of GROND optics with the top panel showing a cut through
   the visual arm, and the bottom panel a cut through the NIR arm.
\label{optlayout}}
\end{figure}

\subsection{Optics}

\subsection{General considerations}

The final choice for the telescope was the MPI-owned 2.2\,m telescope,
operated by ESO on La Silla (Chile); see $\S$\ref{tel}.
This telescope is a f/8.005 Ritchey-Chretien telescope \citep{rit28} 
with a focal length of 17600 mm, on an equatorial fork mount;
the intrinsic image quality is 0\farcs4. 

\begin{figure}[ht]
\includegraphics[width=1.05\columnwidth]{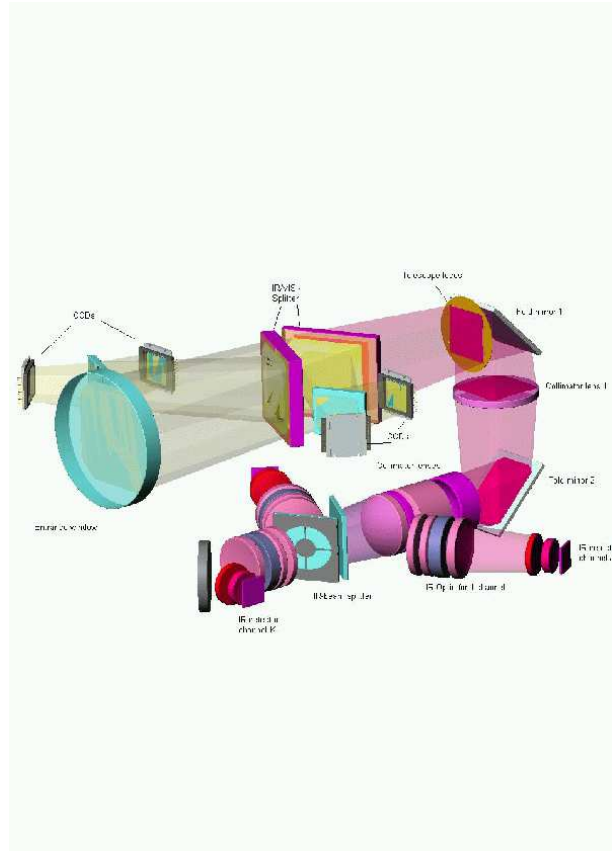}
\caption{3D structure of the visual and NIR beams with most of the components 
labeled.
\label{3Dopt}}
\end{figure}

The prime goal for GROND is to investigate the high-redshift
universe, thus emphasis was put on the NIR channels, and not on the
visual ones. Consequently, it was decided to develop a focal
reducer in the NIR with a 10\amin $\times$ 10\amin\ field-of view
in order to cover GRB error boxes with a typical error radius
of a few arcmin (anticipated to occur about twice per week).
In contrast, in the visual channels, no lens system
was planned and the plate scale provided
by the telescope optics was used. 

The separation of the  different photometric bands has been
achieved using dichroics, whereby the short wavelength part of the light is
always reflected off the dichroic, while the long-wavelength part
passes through. Thus, the first dichroics are used to split off
the visual bands which is done in the converging beam before
the telescope focus.
The relatively short back focal distance of 550 mm implied very tight
constraints on separating the beams and still having enough space for
the detectors.
This, unfortunately, also prohibited using larger-format CCD detectors,
since the consequently larger beams could not be separated further.

After passing a diaphragm at the nominal telescope focal plane, 
the beam is folded by 180\degs\ for space constraints, and fed into
a collimator (Fig. \ref{optlayout}, \ref{3Dopt}). 
The splitting of the three near-infrared 
channels occurs in the parallel beam between collimator and camera objective
of the NIR focal reducer.

One might expect different ghosting effects in the
visual arm (dichroics in converging beam) and the near-IR arm
(parallel beam). While it is true that reflections from the
back surface of the dichroic in the parallel beam produce
ghosts which are not offset relative to the main image,
the effect of these ghosts is negligible. Both the dichroic
as well as the anti-reflection coatings have efficiencies of
about 99.5\% in their transmitting regions. Since back surface
ghosts involve two coating passages, they arrive with $<$0.0025\% 
(corresponding to 11 mag fainter) of the original intensity,
undiscernable in data analysis.

Another typical concern of dichroics is their polarisation dependent
location of the cut wavelengths, in particular since gamma-ray burst
afterglow emission shows variable optical/NIR polarisation. 
We have therefore embarked on a 
dedicated study of this effect prior to the coating of the optics. 
While standard cut wavelengths differ by 7\% (3-4\%) 
for dichroic cubes  (plan-parallel plates) between the 
perpendicular polarisation directions, the final GROND coatings
only have a difference of 1\% (e.g. 15 nm for the $JH$ dichroic coating).  
For a 100\% polarised white light beam
this leads to a \lax 0.1\% change in the transmitted flux, which
in turn implies a \lax 10$^{-3}$\% change in the color terms (see $\S$6.2),
and thus is tolerable.

The entire system has been designed for a working temperature of 80 K in
vacuum, because deviations from the nominal temperature by more than
20 degrees would lead to refractive index changes large enough to
cause noticable image quality degradation,
and we considered it too risky to predict an equilibrium temperature
gradient between the visual and NIR part.
Since the mechanical assembly was done at room temperature, all
optics (and mechanical) system parameters had to be calculated for a
temperature of 80 K. For example,  the focal length of the focal reducer
differs by 3 mm between 293 K and 80 K, compared to a depth of focus
of 25 $\mu$m.  We imposed a  high accuracy
requirement on the mechanical support structure, 
so that an alignment of  the optics system is only
required along the optical  axis.

\subsubsection{The four visual channels}

For the visual channels the incoming 1:8 telescope  beam is split using 4
dichroics (top panel of Fig. \ref{optlayout}). 
The field of view is determined by the telescope focal length
(17.6 m) and the size of the CCDs ($2048^2\,\times\,13.5\mu$m).  
For each CCD the field of view is $5.4\,\times\,5.4$ arcmin$^2$,
and the plate scale is 0\farcs158/pixel.

All four CCDs are backside illuminated E2V devices without anti-blooming 
structures and are operated in inverted (AIMO) mode.
The chips come on  nickel plated ceramic base plates.
Since each of the detectors serves in one and only one filter band,
different sensitisations of the detectors have been applied:
$g'$:  astro-broadband on normal silicon,
$r'$:  mid-band on normal silicon,
$i'$ and $z'$:  basic-NIR on deep-depletion silicon.

The use of dichroics implies that adjacent bands have identical 50\%
transmission wavelengths, making the Sloan filter 
system \cite{fig96} the obvious choice.
Thus, the dichroics were designed such that the combination of their
cut-off wavelengths defines bands identical to the Sloan system, with 
the exception of the $i'$ band. Since in the Sloan system $r'i'z'$ overlap
at their $\sim$70\% transmission values, we decided to compromise the
$i'$ band in favour of standard-width $r'$ and $z'$ bands.

\begin{figure}
\includegraphics[width=0.95\columnwidth, bb=89 105 461 755, clip]{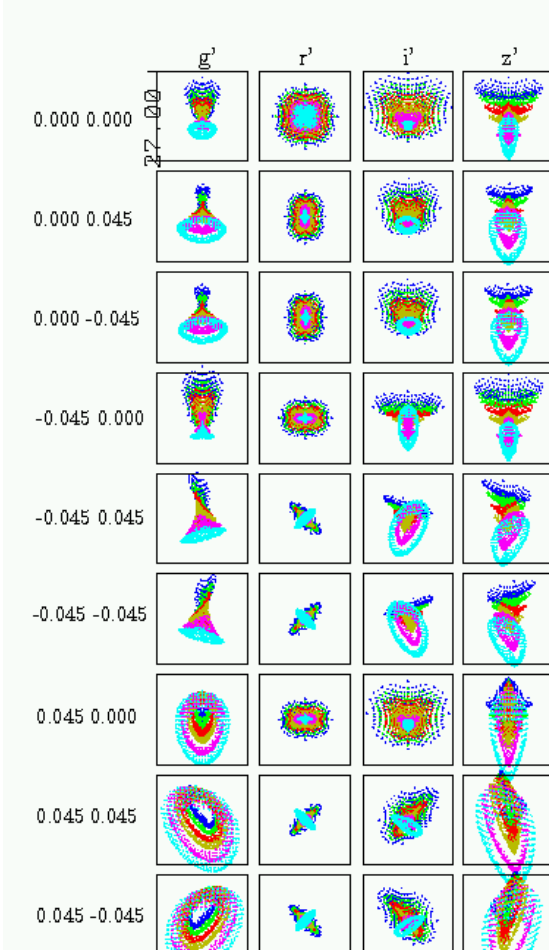}
\caption[VIS spot diagram]{Spot diagram of the four visual channels
   at different off-axis angles (rows): the numbers on the left give x and y
   coordinates in degrees within the field of view, the top row corresponding
   to the optical axis. It shows the geometrical abberations (without
   diffraction) in the focal plane. The wavelength range of each band is
   simulated with three individual wavelengths. The non-symmetrical images
   are due to the wedge shape of the dichroics. Note that the $r'$ channel
   only involves reflections from the dichroics, and therefore is
   symmetric (and thus shows the image quality provided by the three mirrors
   of the telescope).
   Each panel is 27$\mu$m$\times$27$\mu$m, corresponding
   to 2x2 pixels of the detector.
\label{visspot}}
\end{figure}

The telescope  beam is first split by reflecting off the $g'$ and $r'$ bands, 
then the  $i'$ and $z'$ bands.  The long wavelength part of the beam (the
NIR part) is then passed through the primary cold stop in the 
focal plane.
Two further dichroics are placed into each of the two visual beams to separate
$g'$ and $r'$, and  $i'$ and $z'$, respectively.
The dichroic plates are wedge shaped by $\sim$13--19 arcmin 
so that they compensate for the astigmatism of a
plane-parallel plate in the converging 1:8 beam of the telescope \cite{wlp}.
The curvature of the focal plane of the
telescope has a radius of 2205 mm, which corresponds to a focus
difference of 43$\mu$m between the center and edge of the CCD. This is
smaller than the depth of focus of $\sim$60$\mu$m at the shortest
wavelength used, but of course causes astigmatism (see Fig. \ref{visspot}).

Originally, no other optics were planned except for the dichroics.
After first light it became apparent that the $g'$ and $i'$ bands suffered from
single reflections of $r'$ and $z'$ light, respectively, off the backside
of the dichroics, at a strength of about 0.5\% of the incident intensity.
Thus, each of the CCD channels has been equipped with a short-pass
filter to block these single reflections. The backsides of those
3\,mm glass filters have anti-reflection coatings at $>$99.5\% efficiency,
except for the $g'$-band (see below).

Another post-commissioning change has been implemented for the $g'$ band:
originally, no short-wavelength cutoff was foreseen, since we expected
the transmission of the telescope mirrors and the entrance window as well as
the efficiency of the CCD detector to fall off rapidly enough below 
$\sim$400 nm. In fact, the ill-defined cut-off led to a
photometric uncertainty of about 5\% which was deemed unacceptable
(see $\S$\ref{limmag}).
We therefore implemented a short-wavelength edge-filter 
with a cut-off wavelength of 380 nm.

\subsubsection{The near-infrared channels}

The NIR part is designed as a focal reducer system allowing for an optimized
adjustment of the focal length, or equivalently the field of view. 
In this way the required 10\amin\ FOV is imaged onto a
$1024\,\times\,1024$ Rockwell HAWAII-1 array (pixel
size 18.5$\mu$m, plate scale of 0\farcs60/pixel).  
The focal reducer system (5 collimator lenses with
a resulting F$_{\rm coll}$=360\,mm and 6 camera lenses with
F$_{\rm cam}$=129.6\,mm for each NIR channel) has a pupil reduction 
factor of 48.89
and a pupil diameter of 45\,mm.  For the f/8 telescope the
resulting effective focal length is 6336 mm.

A diaphragm is placed in the telescope and
collimator focus, including blocking of the M2 mirror and M2 spider structure. 
The two beam splitters for reflecting off the $J$ and 
$H$ bands are located in the parallel beam of the focal reducer, between 
collimator and the camera objectives (Fig. \ref{optlayout}, \ref{3Dopt}).
In addition, each channel has  a cold pupil stop
near the dichroic, as well as a filter in front of the first camera lens
defining the standard $JHK$ bands (which also prohibit long-wavelengths
single reflections to reach the detector).
The $JHK$ filters differ from canonical NIR filters in their much 
reduced blocking range which  is sufficient for GROND because of the action
of the dichroics. Thus, the mean transmission in the $JHK$ bands 
(including dichroics) is \gax 98\%.

Because of the parallel beam, the dichroics could be manufactured as true plane
parallel plates and mounted in a compact 60\degs\ box which
allows for a space-saving splitting of the beam into the $J,H$ and
$K$ channels. The $K$ channel includes an additional flip mirror for
dithering purposes.

\begin{figure}
\includegraphics[width=0.85\columnwidth, bb=87 110 394 755, clip]{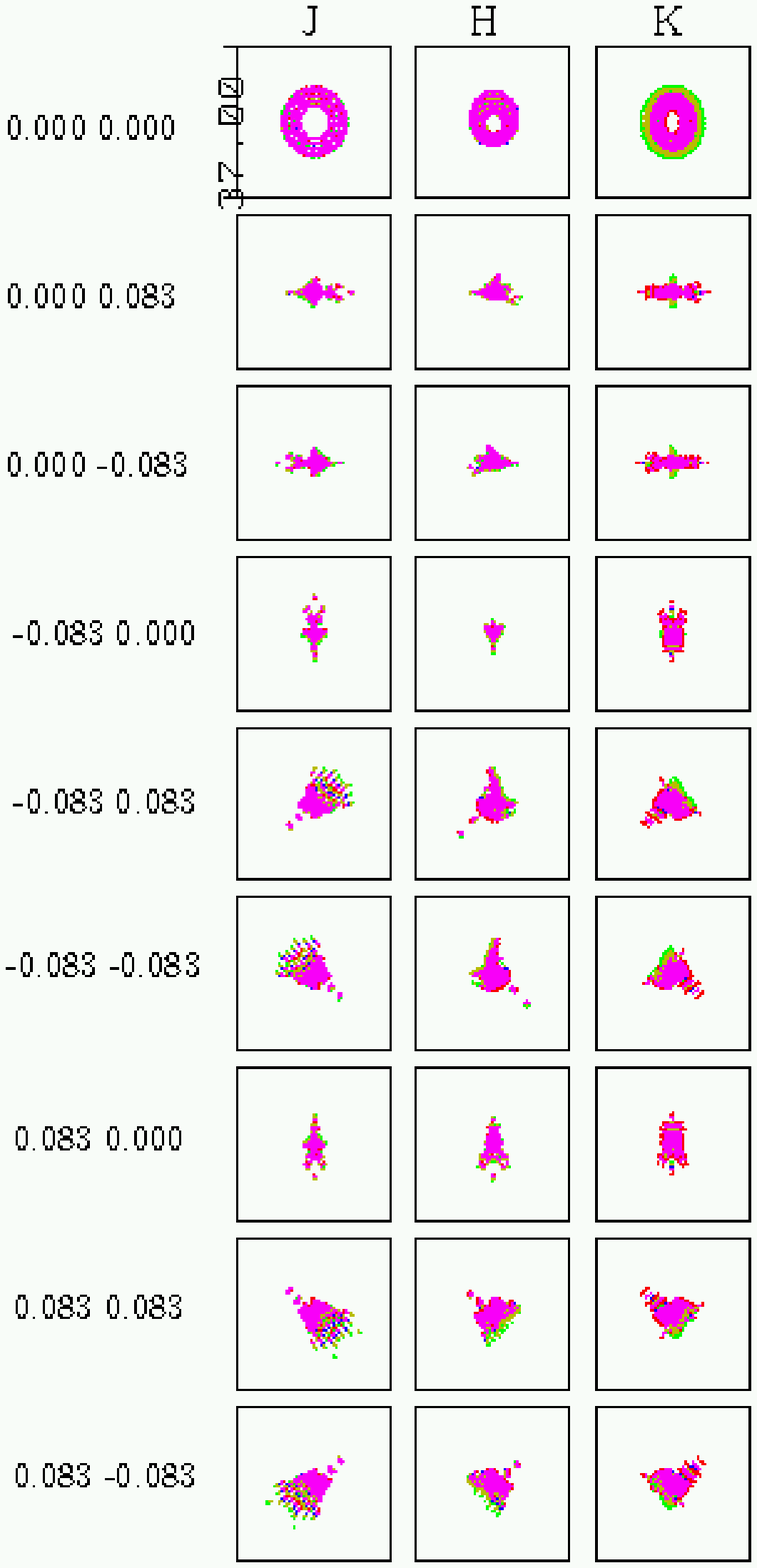}
\caption[IR spot diagram]{Same spot diagram as Fig. \ref{visspot} but for the
   three infrared channels. The focal reducer was designed to optimize the
   image quality (point source shape), but still contains image distortions,
   i.e.  plate scale changes depending on the position in the
   focal plane. These geometrical distortions are
   corrected during the data reduction process (see $\S$\ref{pipeII}).
   Each panel is 37$\mu$m$\times$37$\mu$m, corresponding
   to 2x2 pixels of the detector.
\label{irspot}
}
\end{figure}

The focal reducer system designed  for GROND differs substantially from that
of other optical focal reducer systems (e.g. the Calar Alto Faint Object 
Spectrograph  CAFOS\footnote{www.caha.es/CAHA/Instruments/CAFOS}, 
the Multi Object Spectrogragh for Calar Alto 
MOSCA\footnote{w3.caha.es/CAHA/Instruments/MOSCA}, 
or the Potsdam Multi-Aperture Spectrophotometer PMAS \cite{rot06} at Calar
Alto),  since no optical glasses could be used.  Basically only four 
media are available for transmission in the 900--2400 nm range: 
CaF$_2$, BaF$_2$, OH-free Silica and undoped YAG. Because of its  
transmission
characteristics, YAG is particularly suited for NIR optics (but also difficult
to grow in large dimensions).  Using these four
optical media, and after measuring their refractive indices at
the foreseen operational temperature of 80 K, 
the focal reducer system could be designed with a high performance 
(Fig. \ref{irspot}) using software developed by us (Laux \etal\ 1999). 
The 3D visible and NIR model as well as the optics components
CAD files
were created using the ZEMAX software package\footnote{www.zemax.com}.

\subsection{Cryo-Mechanics}

\subsubsection{Cryostat}

\begin{figure}[ht]
\includegraphics[width=0.99\columnwidth]{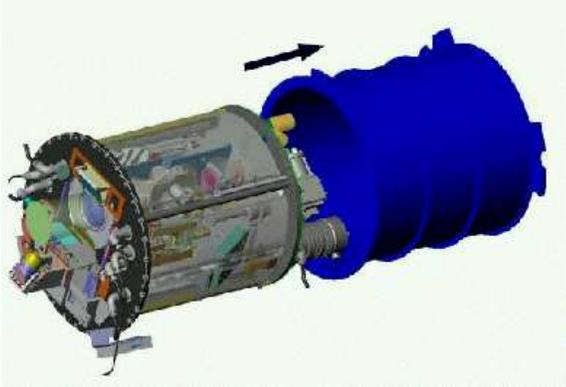}
\caption[Cryostat]{Removing the cylindrical shell of the cryostat allows
easy access to all internal components.
\label{cryostat}
}
\end{figure}

The GROND cryostat (Fig. \ref{cryostat}) consists of the following components:
a) GROND-cylinder, made of titanium, and electron beam welded;
b) the front and back plates made of aluminum;
c) 4 spacer rods;
d) 8 Belleville-spring systems; and
e) O-ring seals (Viton).                              

The components inside the GROND cryostat have different thermal requirements.
In addition,  in the case of single component failures, no sensitive 
components should be exposed to dangerous thermal conditions.
The complete internal structure is mounted on a titanium alloy (TiAlV)
plate that provides thermal insulation. This plate is mounted on the front 
plate of the cryostat,
which accordingly becomes the nominal reference point of the system.
As the different components (front plate, titanium plate, optical bench) have 
different temperatures, the mounting must allow for thermal contractions. 
This is
achieved by using long holes and centering pins to mount the titanium plate. 
The internal components (optical benches) are connected to the back plate of 
the cryostat using a metallic membrane that allows contractions 
only along  the optical axis of the instrument.

The cooling system is based on a two-stage Sumitomo Closed-Cycle Cooler (CCC) 
with a SRDK-408-S cold head which provides 
a cooling power of \gax 35 W  down to 45K at the warm stage, and
\gax 6.3 W down to 10K at the cold stage.
The cold head is mounted using a bellow (to reduce vibrations) on the 
back plate of the vessel.

The first stage of the CCC (higher temperature and higher cooling power) 
is used to cool the optical benches (visual and near-IR) and all 
optics components to 80K (temperature-controlled).
The second stage (lower temperature and lower cooling power) is used to 
cool the near-IR detectors to 65K (optimum temperature to maximise the 
signal-to-noise ratio).
The CCD detectors have an intentionally bad thermal connection to the 
visual optical bench (80\,K). They are stabilised at 165\,K using resistor 
heaters. Thermal connections between the two baseplates and the 
components are realised
using copper flanges, copper ledges and flexible copper lines. 
Galvanic isolation of the cooling system is achieved by isolation foils 
being mounted between the thermal copper flanges and the CCC stages.

To reduce the thermal input, the optical benches are thermally insulated 
from the cryostat. In addition, thermal shields are installed to reduce the 
radiative heat transfer.
The main radiation shield is between the optical benches and the vessel. 
It is thermally insulated from both. The radiation shield is constructed 
of three round plates (two at the front plate, one at the back plate of 
the vessel), four half cylinders around
the spacing bars and six dismountable cylinder segments -- all made of 1.5mm
thick highly reflective aluminum plates.
The cylinder segments are additionally covered with multi-layer insulation
foil.

\begin{figure}[ht]
\includegraphics[width=0.99\columnwidth]{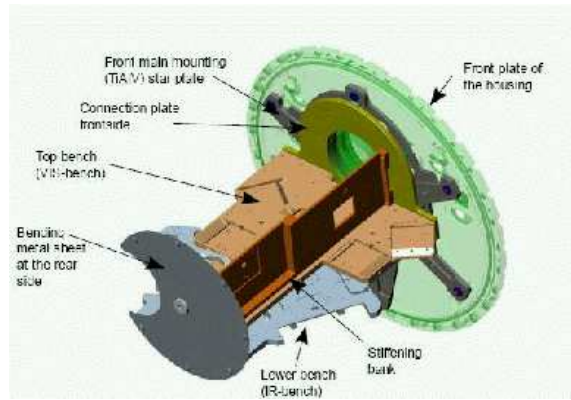}
\caption[GROND structural chassis]{GROND structural chassis with the
two separate optical benches (visual in orange, NIR in blue) and
the main aluminum alloy front plate (grey spider plate).
\label{chassis}
}
\end{figure}

\begin{figure*}[t]
\includegraphics[width=0.6\textwidth]{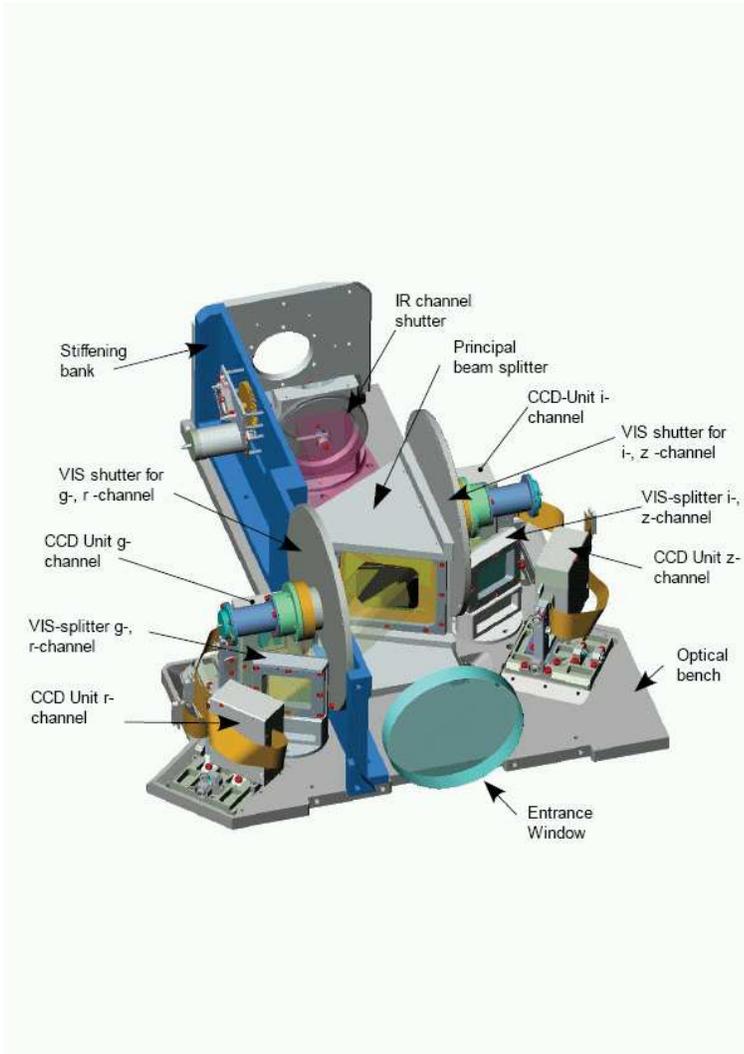}
\caption[Visual baseplate]{Visual baseplate with nearly all optical and 
mechanical components marked.
\label{VISbench}}
\end{figure*}

\subsubsection{Structural Chassis}

The implementation of the optical requirements with the geometrical and 
mechanical constraints such as tight space, minimum weight 
and different 
temperature regions within the main housing resulted in substantial
complexity which in turn required 
most sub-units and their interrelations to be treated simultaneously. 
The development of the optical/mechanical configuration included considerations
such as:
(i) space requirements for the 4 CCD cameras and 3 NIR-detector units,
(ii) use of structural components, such as optical benches, mounting 
  interfaces/connections, and cold connections, 
(iii) assembly and handling procedures, 
(iv) cabling, 
(v) low-temperature induced contractions interacting in 3 dimensions.

\begin{figure*}[ht]
\includegraphics[width=0.6\textwidth]{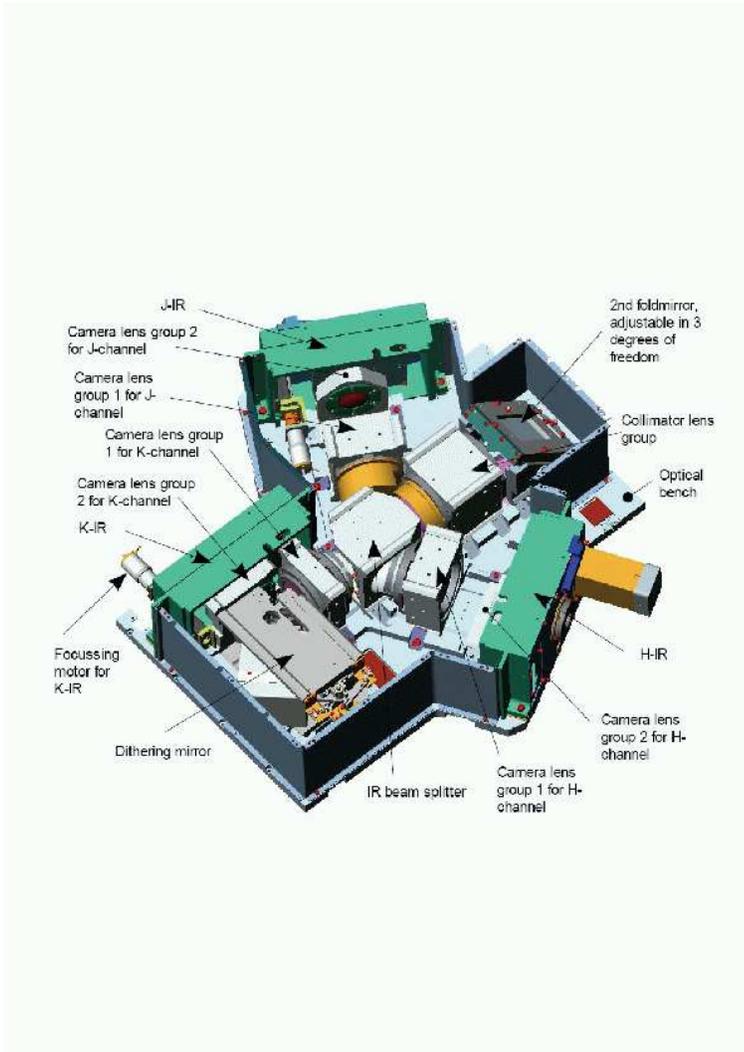}
\caption[Infrared baseplate]{Infrared baseplate with all optical and 
mechanical components.
\label{IRbench}}
\end{figure*}

The final structural chassis (Fig. \ref{chassis}) consists mainly of 
two separate optical benches. The 
top bench (Fig. \ref{VISbench}) contains the optical and mechanical 
groups for the visual light detection, 
in particular:
(1) the entrance light baffle (not shown in Fig. \ref{VISbench});
(2) principal light splitter unit with two dichroics;
(3) small dichroics splitting the $g'$, $r'$ and  $i'$, $z'$ channels, 
   respectively;
(4) two VIS-shutter units for CCD-channels $g'r'$ and  $i'z'$, respectively;
(5) all four CCD-boxes for $g'r'i'z'$;
(6) NIR channel shutter to close the NIR sector completely;
(7) mirror carrier unit for folding mirror 1 (also not shown in 
Fig. \ref{VISbench}).
The lower bench (NIR-area; Fig. \ref{IRbench}) contains the following 
components:
(1) 2nd folding mirror (adjustable in 3 degrees of freedom);
(2) collimator lens unit;
(3) two NIR beam splitters to separate $J$, $H$, and $K$;
(4) one camera lens system (split into two mechanical lens holders, 
   Fig. \ref{optlayout}); 
for each of the 3 NIR channels;
(5) dithering mirror unit for the $K$-channel;
(6) NIR-detector units (motorised for focussing);
(7) Zeolith boxes near the CCC cold stage as water (humidity) sinks
(not shown in Fig. \ref{IRbench}).
The NIR light path is only open at the focal plane, but otherwise
completely closed by a dedicated light-tight housing cooled to 80\,K.
Cabling is done through light traps.

\subsubsection{Lens Mounting}

\begin{figure}[ht]
\includegraphics[width=0.99\columnwidth]{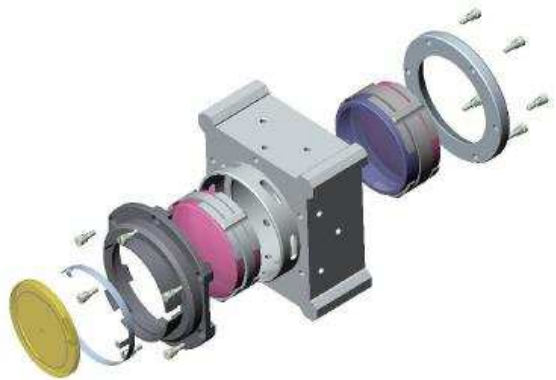}
\caption{Exploded view of the lens mounting block for the 4
collimator lenses in each of the three NIR channels.
\label{Lens}
}
\end{figure}

Within the focal reducer, 23 lenses are maintained
at precise position during the cooling to 80\,K.  
A special spring-based lens mounting design was developed 
(Fig. \ref{Lens})
which keeps the lenses  centered and in the correct axial position. During 
the cooling the radial flexibility of the lenses is kept
small enough to ensure the required centricity.

\subsubsection{IR Unit}

\begin{figure}[ht]
\includegraphics[width=0.6\columnwidth,angle=270]{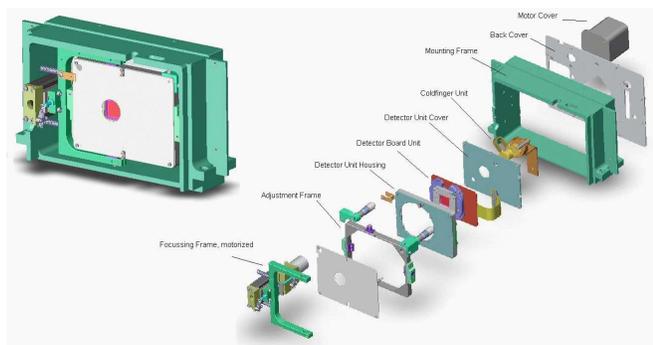}
\caption[IR Detector]{Schematics of the NIR detector unit.
\label{IRUnit}
}
\end{figure}

The tight space constraints also made it necessary to develop a new, much 
smaller detector board (shrinking from originally 220\,mm diameter to a 
square of 50x70 mm). Consequently, the mechanics also had to be
re-designed from scratch and the opportunity was taken to implement two
new features: (i) motorised positioning of the detector along
the optical axis, and (ii) a gimbal mounting to allow manual alignment of
the detector plane perpendicular to the optical axis in both degrees of 
freedom (Fig. \ref{IRUnit}). 

A cold finger on the backside of each
NIR detector is directly connected with the second 
stage of the CCC, thus allowing the NIR detectors to cool 
below the temperature of the baseplate and optics. 
Since the CCC temperature is load 
dependent, a heater is installed on the coldfinger to control the 
temperature and keep variations below $\pm$0.1 K.

\subsubsection{CCD detector unit}

\begin{figure}[t]
\includegraphics[width=0.99\columnwidth]{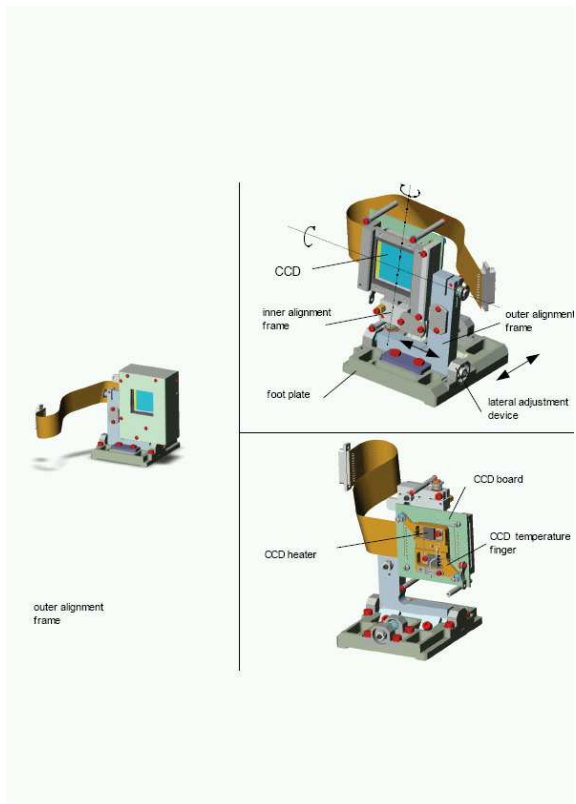}
\caption[CCD detector unit]{Schematics  of the CCD detector unit.
The arrows denote the alignment possibilities.
\label{CCDUnit}}
\end{figure}

The CCD detector units have no motorised focusing stages, 
because the depth of focus is larger.
Still, they can be aligned in 4 degrees of freedom (Fig. \ref{CCDUnit}). 
Since the minimum allowed operational
temperature of the CCDs ($\sim$155\,K) is much higher than the inner cryostat
temperature, the CCDs are equipped with a  heat finger 
and thermally controlled via the PULPO electronics 
(see section \ref{pulpo}) to 165\,K within $\pm$1 K.

\subsection{Cryo-Motors}

There are several motorized mechanisms in the instrument which operate
at 80 K. All motors are from Phytron, but use different gears.

\begin{itemize}
\item \emph{NIR channel Shutter}: This door closes the near-IR arm of
the instrument, so one can take darks in $J$, $H$ and $K$. It is implemented
using a stepper motor. 
\item \emph{Visual Shutters}: 
Due to the constraints of the optical design of the instrument, the
CCD shutters must be cryogenic (as the shutter must not block the
infrared channels).
Due to the tight space constraints, one shutter per CCD was mechanically
impossible; thus two shutters have been implemented, 
one for the $g'r'$ and one for the $i'z'$ channels, respectively.
Each shutter is a thin metal disc with two quadrant shaped
openings at opposite sides that are slightly larger than the beam at 
this position. Thus, each disc has two 'shutter
closed' and two 'shutter open' positions, i.e. 4 possible positions in total
which are 90$^\circ$ apart.
Each shutter is equipped with a stepper motor and a position encoder that 
provides the (absolute) position of the disc. 
The combination of stepper motors
and encoders ensures that movements are fast (with stepper motors one
can 'blindly' move the shutter by 90 degrees) and that the shutter
does not drift over time, resulting in vignetting (i.e. the 4 nominal
positions are tied to encoder positions). This design is still
slightly slower than typical, non-cryogenic shutters, but it is fast
enough (below 1 sec) for typical exposure times of 1 min or more.
The two shutters can be operated independently, thus, different integration
times in the two optical arms are possible.
Shutter movement is triggered by the PULPO shutter control port 
(see section \ref{pulpo}).
\item \emph{Flip mirror zero position}: The flip mirror ($K$-band
channel dithering) can be either in the calibration/alignment position
(i.e. no dithering) or in the normal position (where dithering
is possible by rotating the flip mirror). This function is implemented
using a stepper motor that moves the mirror into either position.
\item \emph{Flip mirror dithering}: This motor is responsible for
the internal  dithering in the $K$-band. Dithering in $K$ as opposed
to the other channels is required due to the high sky brightness.
In order to allow staring at the same sky position in the optical
bands (because of the long CCD read-out times), this internal dither
mirror has been implemented. 
The dither function is implemented using a {\tt CLD}
stepper motor linear drive from Phytron and an encoder. Two fundamental
operational modes are implemented. Either the motor is moved to an absolute
position or the motor moves in either direction 45, 60 or 90 degrees
(i.e. 8, 6 or 4 dither positions). 
\item \emph{IR focus}: Each of the three NIR detectors can be focused 
independently using 3 stepper motors with gears. Each motor allows relative 
movements to the required accuracy of $\pm$25 $\mu$m), over a range of 
$\pm$2\,mm. The gear is
strong enough to prevent any shift of the NIR detector with
positioner in the best-focus position during telescope movements.
 
\end{itemize}

\subsection{Electronics}

The electronics are divided into several parts: 
(i) the main rack which contains the GROND-specific control, 
(ii) the read-out electronics for the CCD and NIR detectors as provided
by ESO.

\subsubsection{Main electronic rack \label{erack}}

\medskip
\centerline{\em Interface Unit and Main control}

This unit forms the control interface between the GROND instrument workstation 
and   the motor control electronics. The main part of this unit is a 
XILINX FPGA, which makes the  electronics more flexible. Each IO of 
this XILINX FPGA is driven by a bidirectional bus driver and additionally 
ESD protected by a low capacitance diode array. The current design uses 
98 pins for input and 66 pins for output purposes, with 32 spare pins.

{\bf Communication interface:}
The communication interface is realised by an onboard RS232 interface 
with two jumpers for speed selection. The nominal operation speed is 
115200 baud. A RS232 server converts this signal to 100baseT for the 
instrument workstation. 
To maximise reliability, a special command structure is used. 
After the detection of the sync byte, the interface unit starts a 
timeout counter. If a command is not completed within 8 ms, this 
command will be rejected and the whole communication unit reset, 
so that this unit can receive new commands immediately.
This protects the system from hanging due to incomplete commands.

{\bf Motor Control:}
The main control contains all electronics necessary for the digital 
control of the motor driver boards and position readout.
All signals which are important for operation of the motor driver 
electronics (clock, boost, direction, reset), are delivered by the 
main control. For all position-switch controlled stepper motors 
(IR-calibration, cold shutter, zero position), the digital circuit 
prevents a motor overrun (i.e. an end switch). This means that as 
long as the stepper motor is in between the two end-switches the direction 
will be determined by the user. If one of the end-switches is reached, the 
motor stops immediately and the direction is automatically reversed.
Likewise, these electronics are responsible for the readout of the 
resolver values and synchronisation with the stepper motors. Three 
resolvers are controlled by the GROND electronics; one for the flip mirror and 
two for the CCD shutters.

An additional offset register can be used to calibrate the start position 
of the flip mirror. The movement is bidirectional and commandable. 
The accuracy of the flip mirror position is about 5.4 arc\-min (4000 steps
per full rotation).
Contrary to the flip mirror, the CCD shutter electronics have only one 
mode (4 $\times$ 90 degrees). 
The shutter is commandable over the PULPO (see section \ref{pulpo}) 
interface, which 
gives a single bit state to the shutter electronics that determines whether
it is open or closed. A commandable offset register for each CCD shutter 
helps the user to adjust the initial open/close position. The accuracy 
of the shutter is also 5.4 arcmin, corresponding to 80 $\mu$m in the
center of the opening.

\medskip
\centerline{\em Motor driver boards for stepper motors}

The motor driver board has two independent driver units. 
These boards are used for driving the NIR focusing stage, 
the cold shutter and the
zero-position motors. Each driver unit has four current modes (boost, 
run, standby, off), which are selected via two status bits. 
The current levels for the modes can be set using variable 
onboard resistors. Additionally, each driver unit has its own temperature 
sensor, which prevents overheating.

\medskip
\centerline{\em Motor driver boards for linear motors}

This board was specially developed for high current with linear motors. 
Linear motors are used for moving the M3 mirror and the main shutter door
in front of the entrance window.
The board works with two limit switches, between
which the motor can move. 
The forward--backward movement of the mirror is possible by 
software-controlled command (from the GROND instrument workstation) or 
manually over a special control box at the telescope's M1 mirror cell.

\medskip
\centerline{\em Resolver Boards}

An AD2S82A represents the main part of the resolver board, which is a variable 
resolver-to-digital converter. The resolution of this converter is 
variable and can be set up to 16 bit, which means that a maximum resolution 
of approximately 20 arcsec is possible. The sine-wave for the LTN 
resolver is produced by the maximum high frequency waveform generator Max038. 
This produces an optimal signal for the resolver. A high speed voltage 
follower is connected at the outlet side to provide a high current 
driver for long cables. Likewise the offset of the sine-wave can be 
adjusted with a voltage buffer. 
All resolver boards used in the GROND electronic rack have the following 
configuration:
resolution:  16 Bit;
peak-to-peak sine wave amplitude: 2 V;
frequency: 10\,kHz;
offset: 0\,V.

Finally, the primary power of all motor power supplies is switched on by a 
zero crossing switch. This prevents a high inrush current on the main 
power line during switch on.

\medskip
\centerline{\em Temperature Control}

The electronic rack has two temperature control systems. 
The temperatures of the NIR detector system is monitored by Lakeshore 
temperature controllers. 
Each of the three NIR detectors has its own Lakeshore controller 
LS331\footnote{www.lakeshore.com/temp/cn/331dn.html}
which in turn has two control loops:
Loop1 with 50\,W maximum heating power is used to warm up and regulate 
parts of the baseplate.
Loop2 with 1\,W maximum heating power controls the detector temperature 
with an accuracy of $\pm$0.1\,deg. The nominal operating temperature of the
GROND NIR detectors is 65\,K. 

For the optical CCDs, 
temperature sensors and controlled heaters  
are directly mounted to the CCD detector backside
and controlled via PULPO (see section \ref{pulpo}). 
Since each PULPO can control only one shutter,
two PULPOs are incorporated in the electronic rack.
The nominal heating power
of the PULPO is 7\,W in each circuit. For safety reasons, the power at 
the CCDs was reduced to 1\,W. A third heater, also controlled by each PULPO,
is mounted on each of the two clock-bias board thermal shields,  
with 7\,W heating power.

\subsubsection{Detector read-out electronics \label{pulpo}}

\noindent{\bf IRACE:}
The readout system for the infrared detectors, the IRACE (Infrared Detector 
High Speed Array Control and Processing Electronics; Meyer \etal\ 1998) 
system, was supplied by ESO.
As read-out mode (\texttt{IMODE}), the double-correlated mode is used,
as it is the only valid mode for science observations.
One change was made to the basic IRACE system, namely 
to read out all three detectors simultaneously and store the data 
into one FITS file. 

\noindent{\bf FIERA:}
The FIERA (Fast Imager Electronic Readout Assembly; Beletic \etal\ 1998) 
system was also provided by ESO.
Implemented changes include
(i) independent read-out for two pairs of CCDs;
(ii) a 3.8\,m instead of the nominal 1.8\,m cable to the front-end electronics
  (which does not increase readout noise).
For GROND, the two-port readout of each detector is used and one FITS 
file with four separate extensions is written after each exposure 
(the two read-out ports are combined into one image before writing to file).
Furthermore, two read-out speeds are typically used: the fast mode
(625 kpixel/s; read-out noise of $\approx$23 e$^-$) in the 4 min and 6 min
observation blocks (see below),
and the slow mode (50 kpixel/s, $\approx$5 e$^-$) in all longer observation 
blocks. The FIERA system includes PULPO \cite{has98}, a separate electronics 
box used to measure and control temperatures (via 3 heaters), measure 
the vacuum in the cryostat (two inputs), and to operate one shutter.

\section{GROND operation scheme}

The GROND operation scheme is designed to be as generic as feasible, 
but primarily suitable for prompt, automatic GRB follow up observations. 
Additionally, it should provide the basic functionality to perform 
observations of any other, non-transient object. 
However, operating a seven channel imager in a single cryostat implies
various observational constraints which should
be considered in the creation of the observation blocks (OBs;
Chavan \etal\ 2000). 
Each observation 
is built by combining exposures (differing in number and length) of the
seven detectors based on an appropriate set of parameters. 
This approach provides the most generic background for science 
observations with the current instrument design.
The most important
parameters are the number of telescope dither positions \texttt{NTD},
the number of telescope pointings  \texttt{NTP},
the number of exposures in the $g'$ and $r'$ band \texttt{NGR},
the number of exposures in the $i'$ and $z'$ band \texttt{NIZ},
the corresponding integration times \texttt{UITGR} and \texttt{UITIZ},
the number of  $K$ mirror dither positions \texttt{NMD},
the number of  $K$ mirror pointings \texttt{NMP},
the number of $JHK$ exposures in a single $K$ mirror position \texttt{NINT},
the number of stacked $JHK$ exposures \texttt{NDIT},
and the $JHK$ integration time \texttt{DIT}.

In order to comply with the GROND science objectives, different default OB 
types are defined for autonomous observations of GRB afterglows. 
The OBs are named after 
the total integration time in the $K$ band, namely 4~min, 8~min, 20~min 
and 40~min OBs, all with 4 different telescope dither positions, 
as well as 6~min, 12~min, 30~min and 60~min OBs with 6 telescope dither 
positions.

\begin{figure*}
\begin{center}
\includegraphics[width=0.5\textwidth, angle=270]{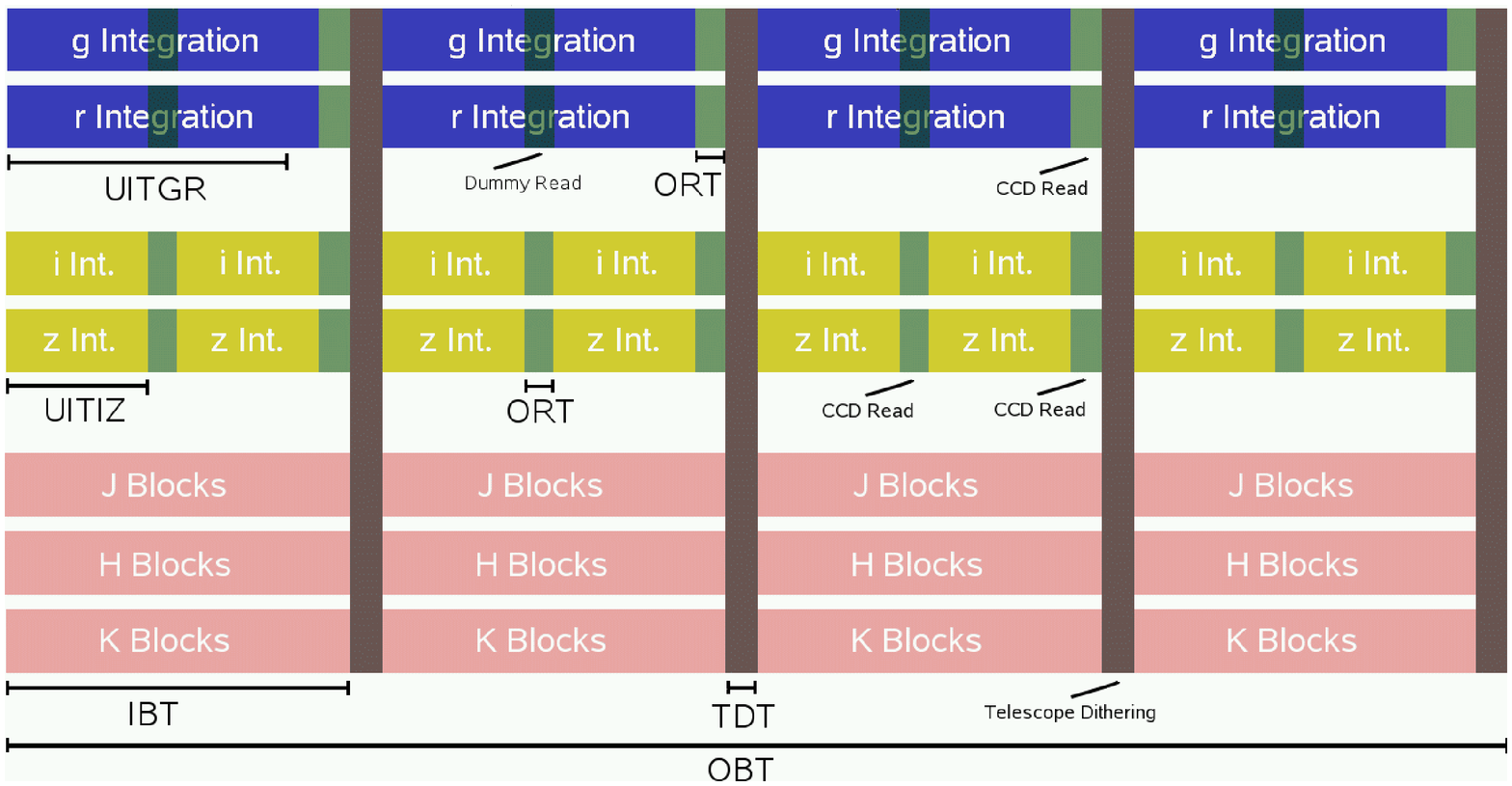}
\medskip
\includegraphics[width=0.45\textwidth, angle=270]{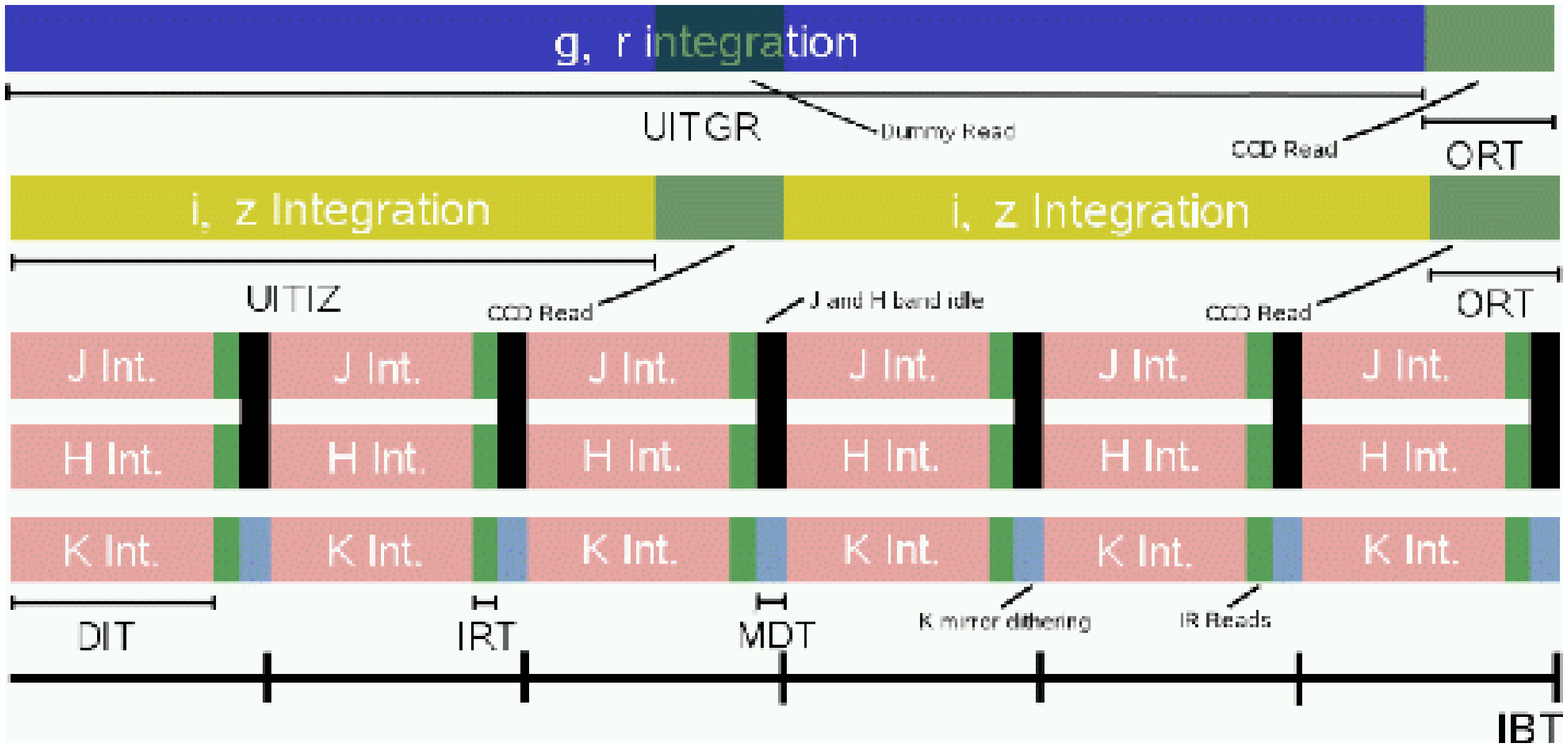}
\caption[Observation block overview]{{\bf Top:}
     Overview of the shortest generic 4~min observation block for
     the FIERA configuration, i.e. the CCD exposures. 
     The read-out (r.o.) of the detectors 
     should not overlap with telescope dithering, as it adds substantial 
     noise to the data. The parameters of the shown 4~min OB 
     are: \texttt{NTD}=4, \texttt{NGR}=1, \texttt{NIZ}=2 and 
     \texttt{OMODE}=0 (625~kpixel/s). With 2 exposures in the $i'z'$ channels
     during one exposure in the $g'r'$ channels, the data in $g'r'$ are not
     read out, but an (empty) FITS-file is created (so-called dummy read-out).
     Descriptions in the
     image apply to all equally colored regions (\texttt{ORT} $\equiv$
     CCD readout time, \texttt{TDT} $\equiv$ telescope dithering time, 
     \texttt{IBT} $\equiv$ duration of a single infrared block,
     and \texttt{OBT} $\equiv$  duration of 
     the whole observation block).
   {\bf Bottom:} 
   Substructure of the observation blocks for the infrared exposures. 
   The read out of infrared  detectors has to be separated from 
   the K$_S$ mirror dithering. 
   The parameters underlying the shown substructure of infrared 
   blocks are: \texttt{NMD}=6, \texttt{NINT}=1 and \texttt{IMODE}=0 
   (double correlated readout).
   (\texttt{IRT} $\equiv$ infrared detectors read out
   time, \texttt{MDT}  $\equiv$  $K$ mirror dither time.)}
\label{fig:GRONDcam}
\end{center}
\end{figure*}

All parameters for GROND observations 
can be adapted for use in non-standard OBs, within their constraints
as described above.
However, all bands and their readout have to be synchronised for telescope 
dithering, which means that the integration time of the optical bands 
must be adapted to the parameters of the infrared observations (and/or 
vice versa). 
For non-standard OBs, this remains the responsibility of the user. 

By default, one standard OB consists of four or six telescope 
dither positions: \texttt{NTD=NTP=4 or 6}, each with six  $K$ 
band mirror dither positions (\texttt{NMD=NMP=6}). A total of
\texttt{NINT}=1, 2, 5 and 10 single $J$, $H$ and $K$ integrations 
are possible at each $K$ mirror dither position.

A generic break-down of the shortest 4~min OB and the underlying structure 
in a single telecope dither position can be seen in Fig. \ref{fig:GRONDcam}. 
Schemes for the other default OBs differ in the number of $J$, $H$ and 
$K$ integrations (\texttt{NINT}) and telescope ditherings (\texttt{NTD}), 
and the CCD detector read-out mode (\texttt{OMODE}).

\section{Software}

\subsection{Instrument software \label{inssw}}

The GROND instrument software is based on the standard ESO/VLT  software 
environment\footnote{www.eso.org/projects/vlt/sw-dev}.
The CCD detectors are handled as a normal array of 4 chips,
each with 2-port read-out.
The single FIERA controller drives all 4 chips. The two shutters
are operated in parallel. In order to integrate longer in the
bluer ($g'r'$) channels, two special read-out modes were introduced:
one that drives only the $i'z'$ CCDs (i.e. one can read out only these two
chips while leaving the image in the other two intact) and a second read-out 
where the $g'r'$ chips are not wiped/reset at the start of the exposure.
When combined, these two readout modes
allow a special operation mode where one takes one $g'r'$ image for
every two $i'z'$ frames. 
Thus, in the bluer channels the
efficiency is significantly improved due to the additional exposure time.
There is a resulting trade off between the dark time (between reset/wipe
and read-out) and the exposure time in the $g'r'$ channels. The latter
cannot be precisely known because of the not exactly predictable 
overhead of a FIERA exposure. 
As the dark current is negligible in the
CCDs, this trade off is justified. For very short exposures of very bright
targets a fast read-out mode is also provided, which reduces the
read-out time from 46\,s  to 4.4\,s.

The near-IR chips are operated normally. The 3 chips with 4 amplifiers
each are treated as 12 video channels and are stored in a single frame.
This implies that all three chips must be read out together. As a
consequence, the efficiency of the $JH$ channels are slightly reduced:
This is because the integration time for faint targets is limited by the 
sky brightness
in the $K$-band to $\sim$10 seconds while the other two bands could,
in principle, integrate 
longer. This results in a 2 second overhead.
In addition, the $JH$ chips must also wait for the dither mirror
movements resulting in an additional 0.5\,sec overhead for each mirror 
movement.
There is no additional read-out noise due to the larger number of $JH$ reads
since the total noise is dominated by Poisson noise of the sky 
background.

The two arms (optical and near-IR) are operated in semi-parallel. Both
detector systems are started independently consecutively (within milliseconds)
and the requested number of frames
are taken with as little dead time as possible. In general
the two detector systems
are {\em not} synchronised -- except after telescope slews.

The instrument control software has no provisions to fill the time
available between telescope movements. It is the user's responsibility
to request the optimal number of exposures. In order to ease this task, a
few optimal configurations are offered.

Relatively frequent telescope offsets are required 
to be able to determine the sky variations in the $J$ and $H$ band
(the $K$-band is handled by the
built-in dither mirror). in order to increase efficiency and to simplify usage,
the dither pattern is a predefined circle -- the user only defines the
number of dither positions. GROND  uses so-called 'combined offsets',
i.e. guiding continues without operator interaction after a telescope
offset. The operator is only required to select the guide star
at the beginning of an OB.

Observations, both in visitor and service mode, are done through OBs. 
The user can use the normal p2pp tool (P2PP Manual 2007\footnote{see 
www.eso.org/observing/p2pp/P2PP-tool.html\#Manual})
like for any ESO instrument
(despite GROND not being an official ESO instrument). 
In the case of GRB observations, OBs are generated in real time by a
special process. The input parameters are provided by the GROND pipeline
host (see next section), which is translated into a standard OB. 
To reduce overheads, the
instrument workstation always provides the current telescope pointing to
the pipeline machine. Thus, the pipeline may conclude that the current
pointing is 'good enough' (i.e. the aim coordinate and error circle fall 
within the field-of-view of the optical CCDs) and skip the telescope preset. 
In this case
guiding will continue without operator interaction, even when executing
several OBs consecutively (guiding does not stop at the end of an OB).
At the start of each OB, the instrument is focused by moving M2
(see  $\S$\ref{foc}. As GROND
is fine-tuned for GRB follow-up where speed is crucial, no manual focusing
is foreseen before a GRB observation starts -- but the necessary 
templates are available for standard observing runs.

To make data analysis easier, all images produced by GROND contain
important information in the FITS headers. Thus, the images can be fed into 
standard software such as IRAF as the required header cards describing
the bias/trim/overscan regions and amplifiers are included. The visual
channels are also identified by the header ($g'r'i'z'$); the infrared channels
are in one image (see above). Obviously, all other relevant parameters 
(gain, read-out noise, etc.) are also included for each chip in the header.

Crude astrometry (WCS cards) is also included in each image. In the
optical the astrometry is correct, apart from a shift introduced by
the pointing of the telescope. The astrometry is correct for all
chips (i.e. individual chip shifts, rotations, etc. are included).
As there are no significant distortions in the optical channels, the
astrometry is very accurate (0\farcs2) -- once the pointing origin is fixed.
In the near-IR channels, the astrometry does {\em not} include the
focal plane distortions (which is significant due to the focal reducer).
Furthermore, the near-IR detector system produces a single image so
an extra processing step is required to slice the image into the
three bands and to extract the necessary information from the header.
The header cards also take into account the $K$-band dithering mirror
position.

\subsection{The GROND Pipeline I: Observation scheduling}

\begin{table*}[ht]
\caption[GROND pipeline duties]{The duties of the system and the 
GRB analysis layers of the GP. \label{sysana}}
\begin{tabular}{ll}
\hline
\hline
System - Observation Control Layer & GRB Analysis Layer \\
\hline
Receiving GRB alerts &  Pre-processing the images\\
Deciding whether to observe the target & Photometric analysis of 7-band data\\
Calculating visibility of the target &  Constructing the SED of the objects \\
Scheduling of the observations & Identifying the GRB afterglow \\
Triggering/continuing/stopping observations & Determining the photometric 
redshift \\
Providing web-interface for user interaction & Evaluating the accuracy of the redshift \\
\hline
\end{tabular}
\end{table*}

The GROND Pipeline (GP) system  is a software package designed and 
written specifically for GROND. Its prime objective is to schedule
rapid GRB afterglow observations (part I; described in this section)
and determine the redshift as quickly as possible (part II; described in the
$\S$\ref{pipeII}). All the components of the GP are deployed on a PC
-- the so-called pipeline machine -- which is 
located in the telescope building.

The coordinates of a GRB are distributed to the world through the Gamma-ray 
bursts Coordinates Network (GCN) in a few seconds (Barthelmy \etal\ 2000).
GROND reacts to GCN notices.
The architecture of the GP is based on an asynchronous framework to 
provide speed and the degree of freedom necessary to apply different 
analysis strategies. In the context of the GP system, asynchrony means 
that all tasks of Tab. \ref{sysana} 
are distributed among different processes which do not run 
sequentially but asynchronously.

The GP mainly consists of two layers, the system layer and the GRB analysis 
layer (see Tab. \ref{sysana}). The system layer consists of the 
processes that receive the GRB alerts, decide whether to follow that 
burst or not, schedule and re-schedule observations and conduct the 
observations by initiating, continuing, interrupting or ending them. 
Furthermore, the main system process controls all processes including 
the analysis processes, and coordinates the interprocess communication.
The GRB analysis layer contains pre-processing of the images, photometric 
analysis, identifying the GRB afterglow, spectral energy distribution (SED) 
analysis and photometric redshift determination.

The details of the processes are described in a separate paper
(Yolda\c{s} \etal; in prep.).

\subsubsection{Receiving GRB Alerts}

When a GRB alert comes in via the GCN socket 
connection\footnote{see gcn.gsfc.nasa.gov/gcn\_describe.html},
the main process extracts all the information 
from the packet by parsing it according to its type. It first checks 
whether it is a packet for an existing GRB or a new target 
for the system. Then the decision tree splits according to packet types.
Based on 
this information and the pre-defined user decision 
on the target type, the system decides whether to follow-up.
For a new target, the visibility of the target is calculated independent 
of the autonomous decision of the software to follow or not.
The visibility calculations utilize {\em skycalc}, a 
C program written by John 
Thorstensen\footnote{see www.eso.org/observing/skycalc\_notes.html}, 
with a python wrapper. The visibility of the object is normally calculated 
for the given/current time or the upcoming night.
If it is observable 
and does  not conflict with other checks (e.g. Moon distance),
then it is scheduled for observation. 
Scheduling means that the main process prepares an observation plan 
of the target (which typically will contain several OBs, possibly also
of different lengths), satisfying the 
above-defined criteria, and tries to 
schedule it with the other observation plans, if any exist. 

\subsubsection{Scheduling of Observations}

The scheduling of observations for GROND needs to be fully automated like 
the other parts of the system, but at the same time it should allow 
users to modify, add, and delete the scheduled observations. Furthermore, 
the automatically scheduled observations may be deleted by the system, 
as a result of a ``Retraction'' or a later ``Ignore'' decision derived from 
the GCN packets for that target. 

Some of the other robotic or automated telescope systems, e.g. the
Liverpool telescope \citep{gmg06} 
use ``just in time'' ({\it jit}) scheduling which is 
based on choosing the observation to be conducted instantly at every 
time rather than scheduling a set of observations for a whole night. 
{\it jit} scheduling is not suitable for GROND because i) ideally,
all GRBs should be observed that occur on the same night and that are visible;
however, {\it jit} scheduling does not allow the system to foresee 
the night and hence arrange the observation durations accordingly; 
ii) the user should be able to interact with and modify the schedule,
e.g. to implement other priorities based on other information (short
duration GRB, or GRB at high redshift);
iii) GROND shares the telescope with two other instruments for which
the observers/operators want to know the schedule of the upcoming hours.
Therefore, 
we developed our own scheduling system that fulfills all the requirements 
of GROND.
The resolution of conflicts between competing observation plans is done 
based on a priority scheme. By definition, at the times when the planned 
observations need to be re-scheduled, the observation plan with higher 
priority can partially or totally override another observation plan 
with lower priority. For GP, the latest burst has the highest priority. 
However, the users can also change the priority of any observation.

\subsubsection{Automating the GRB Afterglow Observations}

When the start time of an observation is confirmed, the GP system triggers the 
Rapid Response Mode 
(RRM\footnote{see www.eso.org/observing/p2pp/rrm.html}) 
by sending the coordinates of the target and the name 
of the observation block to be used to a dedicated directory. 
The RRM procedure has been widely used for the VLT, 
and was implemented at the La Silla site for GROND operations 
on the 2.2~m telescope. 
A RRM acquisition template ends any ongoing observation,
slews the telescope,  and sets up GROND.
After the RRM trigger, the next observation blocks are sent 
directly via the Instrument Work Station (IWS), rather than the dedicated
RRM directory.
Because the GP system does not have direct control over the GROND instrument,
commands and OBs are relayed through the IWS using a specifically designed 
communication protocol.
Once the RRM is accepted, 
the main process of the GP system creates a  GRB process called 
GRBServer. This GRB process controls several sub-processes,
responsible for the analysis of all the 7-channel images obtained for that 
observation, created by typically several subsequent OBs.

\subsection{The GROND Pipeline II: Analysing GRB Afterglow Observations\label{pipeII}}

The analysis of GROND GRB afterglow observations are conducted on the fly, 
managed and controlled by the GP system. 
When an observation is executed, images appear in a certain directory,
as the data handling system (DHS) of ESO pulls them from the IWS and stores 
them in a data directory on the pipeline machine. The analysis of each 
image starts immediately after it appears in the data directory.

The steps of the data analysis  depend on the structure of the OB and the 
choice of the analysis strategy by the users. In total 22 combinations of data 
analysis strategies utilizing different processes chosen from a total of 12 
are available.
The aim is to  construct as quickly as possible a spectral energy 
distribution (SED) of the target object 
to determine its photometric redshift.

\begin{figure}[t]
\includegraphics[width=5.5cm,angle=270]{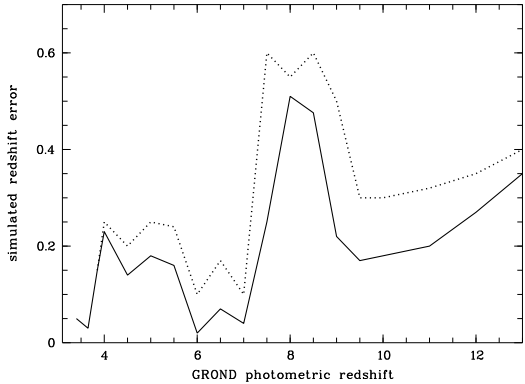}
\caption[GROND photometric redshift error]{The error in GROND's 
photometric redshift determination
of GRB afterglows is expected to be smaller
than $\pm$0.3  for most of the redshift range. Only
for redshifts where the Ly break falls between the $z'$ and $J$ bands,
i.e. $z \approx 8$, does
the accuracy disimprove. The solid line shows the simulated error
in redshift when assuming $\pm$0.5 mag relative photometric accuracy
between the 7 bands. The relatively good performance around $z\sim6$ is due to 
the narrow $i'$ band in GROND.
The dotted line shows the (larger) error
when in addition one assumes an intrinsic GRB host extinction 
of $A_{\rm V}$=1 which is not accounted for in the fit. 
For low redshifts there is 
hardly any difference since the NIR bands provide a precise lever arm.
\label{deltaz}}
\end{figure}

In all strategies, the single band images are corrected for the 
dark current and bias, introduced by the detector and detector electronics, 
and the multiplicative 
effects, i.e. the pixel-to-pixel sensitivity and the illumination 
variations across the image. 
The remaining steps, including i) correction for the geometrical
distortion introduced 
by the focal reducer lenses in front of the infrared detectors,  
ii) sky subtraction for the infrared images, iii) shifting and adding of the 
images obtained at different telescope positions (dithered), vary depending 
on the strategy. Overall,  different strategies can be grouped into two, 
depending on when the astrometric and photometric analyses are conducted. 
The astrometry and photometry can be conducted either only on the images 
acquired at the same telescope position, or at the end of an OB. 
All the analysis processes until the identification of the GRB afterglow 
utilize Pyraf/IRAF libraries\footnote{IRAF (see iraf.noao.edu) is 
a data reduction and analysis software package of NOAO, and Pyraf 
(see www.stsci.edu/resources/software\_hardware/pyraf) is a 
Python wrapper for IRAF, provided by the Space Telescope Science Institute.}. 
Astrometry is done by matching the objects detected in the images with those 
that are in the optical or infrared catalogs, namely USNO A-2, USNO B1, DENIS,
2MASS, NOMAD and GSC22 which are downloaded via the
internet\footnote{vizier.u-strasbg.fr/viz-bin/VizieR}. 

The objects found in each of the 7 bands are matched based on their 
world coordinates, after the photometric analysis. The result of the 
match gives us SEDs of all the objects detected in an OB. The identification 
of the GRB afterglow among the many objects detected in the image is based 
on a figure-of-merit approach. Marks are given to each object as a result 
of several tests, namely whether 
i) the object is inside the area given by the gamma/X-ray/optical 
position uncertainty distributed by GCN packets, 
ii)  the coordinates of the object match with any object in the 
catalogs, 
iii) the magnitudes of the object show variability in time, and 
iv) the colors of the object resemble those of other GRB 
optical/infrared afterglows (Rhoads 2001). 

\begin{figure}[t]
\includegraphics[width=0.73\columnwidth,angle=270]{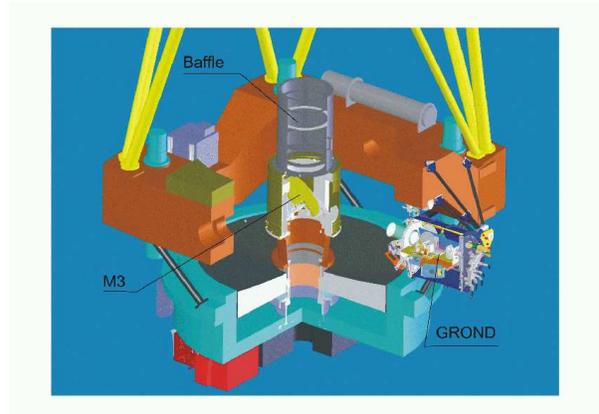}
\caption[GROND at the telescope]{Cut-out view of GROND at the telescope
with the new M3 within the light baffle.
\label{gat}}
\end{figure}

Before the SEDs of the best candidates are analysed by the publicly available 
HyperZ code (Bolzonella \etal\ 2000),
they are first tested for a Ly break.
A single and a broken power-law model are fit
to the SED of each object, and three tests are applied on the results of these 
fits in order to decide whether 
the Ly break is covered by the data.
The redshift range that can be observed by GROND is $z \sim$ 3.5 -- 13. 
The HyperZ tests we conducted with simulated and real afterglow data 
showed that HyperZ is able to determine the redshift with an accuracy 
of $\Delta$$z$ $\sim$ 0.3 -- 0.5 (see Fig. \ref{deltaz}), if the data 
cover the Lyman feature. 

\section{GROND at the 2.2m telescope \label{tel}}

The 2.2\,m MPI/ESO telescope is the obvious first choice for GROND.
In order to detect a large fraction of GRBs, a 2\,m class telescope
is a minimal requirement.
Working within the ESO environment would allow follow-up observations
at the VLT with rather simple procedures as well as providing the
(European) GRB community with unique data. Moreover, it was expected
that the pressure on the  2.2\,m MPG/ESO telescope might decrease
once VST and VISTA become operational.

\subsection{M3}

In order to produce the smallest possible impact on the operation of the
other two instruments on the 2.2\,m telescope, the
Wide Field Imager 
(WFI\footnote{www.ls.eso.org/lasilla/sciops/2p2/E2p2M/WFI/}) and
the fiber-fed Echelle spectrograph 
FEROS\footnote{www.ls.eso.org/ lasilla/sciops/2p2/E2p2M/FEROS/},
we designed a movable M3 mirror which, in case of a GRB alert,
is folded in and
reflects the light off towards the side of the telescope
(Coud\'{e}-like focus) where GROND is permanently mounted.
This leaves the common-user instruments WFI  and FEROS
unaffected at the Cassegrain focus of the telescope. 
Since  GROND observations start as soon as possible after a GRB alert,
the switching mechanism had to be fast, rigid, and 
reproducible in its alignment.

The M3 mirror itself has an elliptical shape with a dimension of
$492\,\times\,324 \,\times\,60$ mm$^3$. The dimensions of the mirror
are large enough to guarantee that additional off-axis light is available 
for a guiding camera sitting on
top of the GROND vessel (see  $\S$\ref{guide}). 
The mirror consists of CeSIC 
(carbon-fiber reinforced silicon carbide).
The advantage of this material for our application is its low  density 
and high rigidity so that the
weight of the M3 mirror is only 8.2 kg; the mass of the total movable part  is
23.5 kg. 
Also, this material allows the mirror and support structure to be
thin enough that no obstruction occurs when in the up-right position for 
WFI/FEROS observations.

The complete M3 mirror system (Fig. \ref{gat})
can be switched to the
GROND position within about 20 seconds. The accuracy of the alignment with the
optical axis was tested several times and found to be better than 10\asec.

\subsection{Pointing Model}

Due to the very asymmetric position of the GROND cryostat vessel
on one side and the electronics on the opposite side of the telescope
centerpiece, a new pointing model had to be created using the TPOINT
package \cite{wal07}.
In addition to the components used for the Cassegrain instrumentation,
two new components (HCEC and HCES, both unrelated to the GROND assembly,
but correcting an earlier omission)
were added in modelling the offsets of the 100 stars which had
been measured at different azimuth and elevation angles.
The effect of GROND on the pointing model is absorbed by those
components which had already been implemented (most notably the fork
flexure and the tube flexure).
The resulting pointing model leaves an rms scatter of 5\asec\ for both,
GROND and WFI (without
the compromise of ensuring a good WFI/FEROS pointing it would be \lax2\asec). 

\subsection{Pointing direction dependent Focus \label{foc}}

Due to the location of the GROND camera on the centerpiece,
the bending of the telescope at different sky positions
introduces a variable focus change. This focus change only depends on 
the declination,
since bending in the plane of
the telescope fork only leads to a lateral displacement of the
beam away from the optical axis, but not a focus change.
The focus changes correspond to about 60 $\micron$ of M2 movement
between zenith and 70\degr\ zenith distance, as
compared to the 66 $\micron$
focus change induced by a 1 degree temperature change.
Switching in/out the M3 mirror reproduces the focus to within $\pm$8$\mu$m.
Automatic re-focusing has been implemented at the template level
and thus is executed at each start of an OB.
In addition to the temperature T, the focus formula 
(F= const. - 66*T - 80* cos(DD); in units of $\mu$m) accounts for
the telescope position in the North-South direction (DD is the 
declination difference relative to zenith). 

\subsection{Guiding \label{guide}}

\begin{figure}[ht]
\includegraphics[width=.65\columnwidth,bb=110 350 335 670,clip]{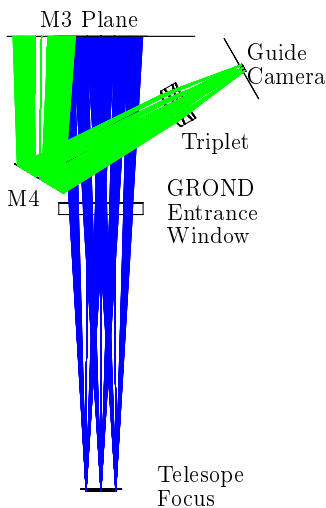}
\caption[GUideCam]{Layout of the guide camera, with the M4 pick-off mirror 
next to the entrance window.
\label{guidecam}}
\end{figure}

In accordance with ESO standards it was mandatory for GROND to be equipped 
with a guide camera.
In order to ease integration with the existing telescope control system.
the ``New Generation Technical CCD'' 
(NGTCCD\footnote{ftp://ftp.eso.org/pub/vlt/vlt/pub/releases/JAN2006/vol-4/VLT-MAN-ESO-17240-3558.pdf})
system of  ESO
with a 1K$\times$1K E2V chip was implemented. In view of the 
semi-robotic operation, a field of view large enough to always contain 
a star brighter than $V \sim 15$ mag was deemed necessary.
Therefore, the f/8 beam of the telescope was re-imaged to yield a
0\farcs33/pix plate scale, and a FOV of 5 arcmin across.
The NGTCCD was placed outside of the main GROND vessel, and is fed by 
a pick-off mirror (M4) next to the entrance window of GROND 
(Fig. \ref{guidecam}).
A folding mirror (not shown in Fig. \ref{guidecam}) is introduced between the
triplet lens system and the detector to ease mounting on the GROND vessel. 
The M4 pick-up mirror
is a weakly spherical mirror (80 km radius of curvature) 
which is slanted by 2\degr\ relative to
a standard symmetric use, thus minimizing astigmatism. The triplet
lens system adapts the plate scale to the chosen detector 
(pixel size and number).
The guide camera's FOV is located 23\amin\ south of the main GROND
imaging FOV.

\begin{figure*}[th]
\includegraphics[width=80mm]{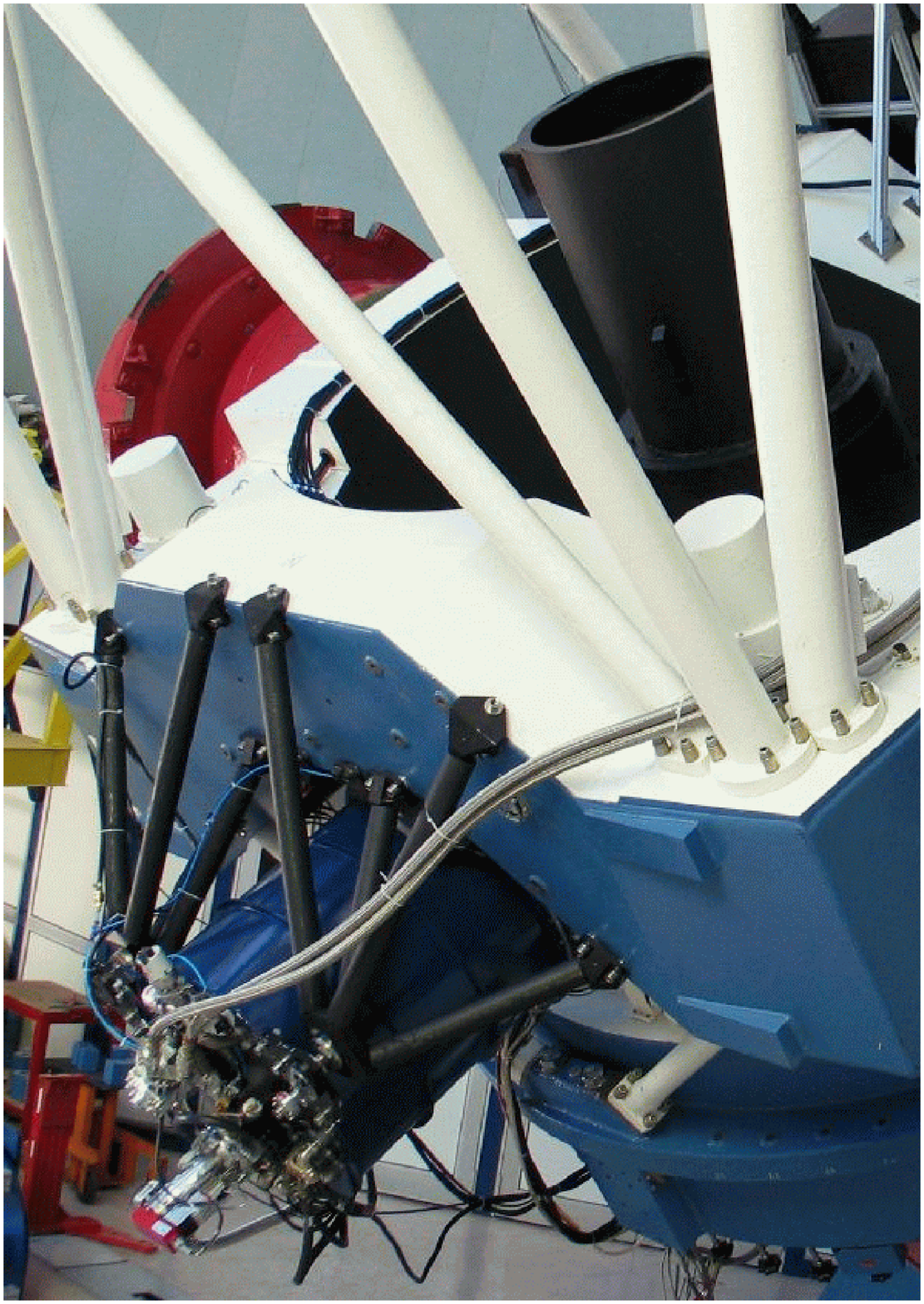}
\includegraphics[width=83mm]{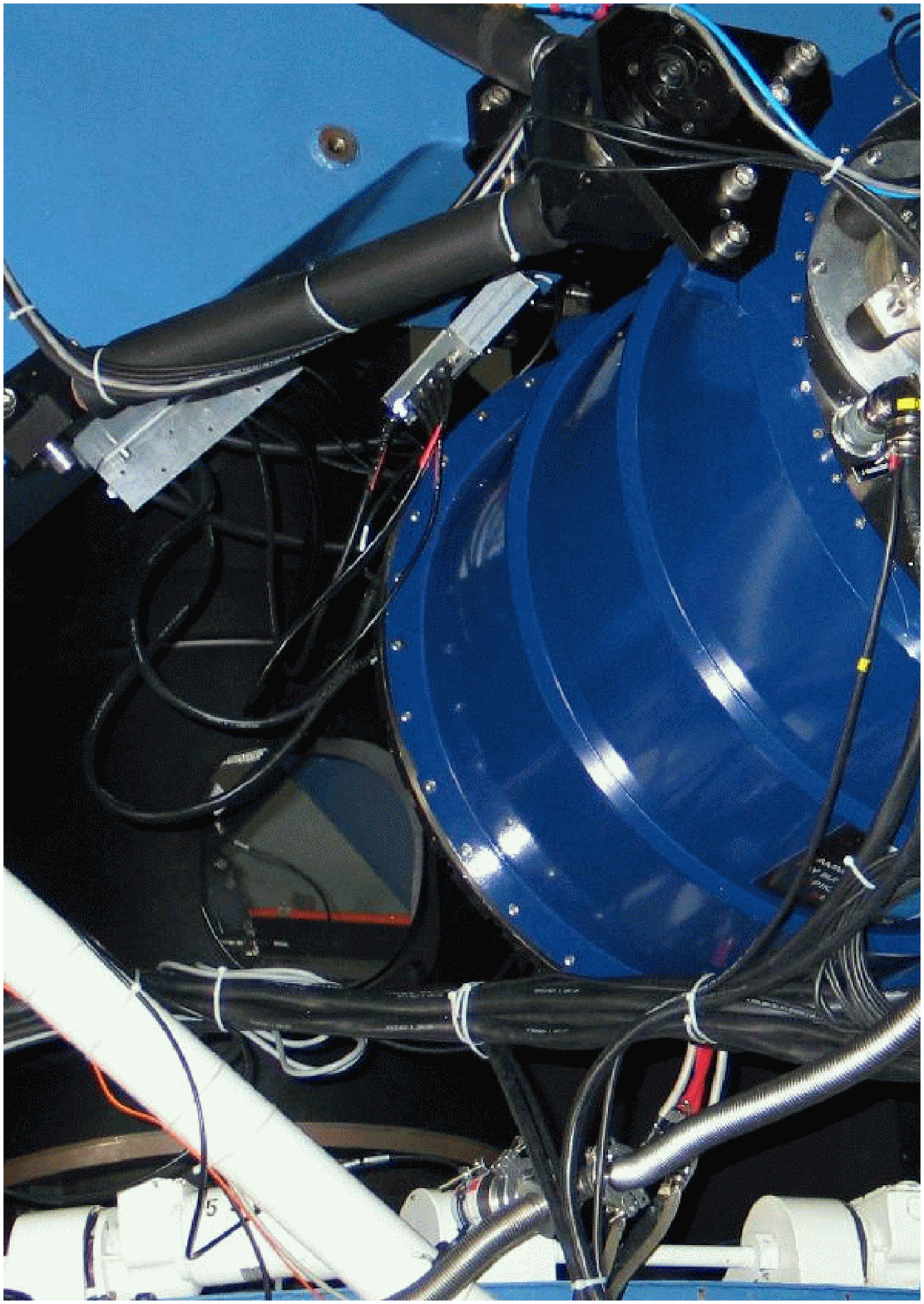}
\caption[GROND at the telescope]{The GROND instrument at 
the 2.2\,m telescope. 
Left: Grand view of the telescope, GROND to the lower left, the light baffle
on top of M3 in the middle, and the main electronics rack at the top.
Right: Blow-up of the GROND vessel and its various connections.
The backside of the cryostat contains the CCC, the turbo-molecular pump, as
well as a small emergency pump for rare cases of power outages.
\label{GAT}}
\end{figure*}

\section{Calibration and performance}

\begin{figure}[ht]
\vspace{-0.2cm}
\includegraphics[width=1.06\columnwidth]{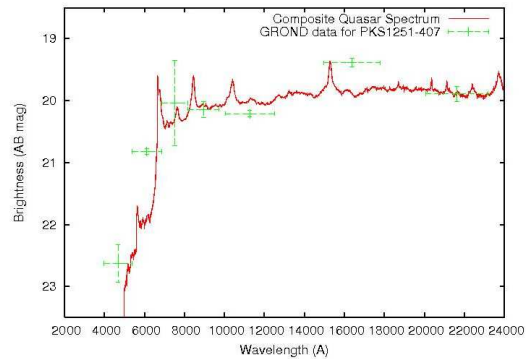}
\vspace{-0.7cm}
\caption[PKS1251]{First scientific demonstration result from April 30, 2007:
green crosses are the GROND camera photometric estimates as derived from 
a 8 min observation block of the quasar PKS 1251-407. The derived
photometric redshift is $z$=4.44, compared to the spectroscopic redshift of 
$z$=4.46. The red line is the composite quasar spectrum from 
Francis \etal\ (1991).
\label{pks}}
\end{figure}

GROND was mounted at the 2.2\,m telescope in April 2007 (Fig. \ref{GAT}), 
and saw first light on the sky on April 30 (Fig. \ref{pks}). 
GROND has been running smoothly since. Particularly
noteworthy are three effects: (i) there has been no need for continuous
pumping; the CCC acts as a cryo-pump, and keeps the vacuum stable
at $\approx 5\times 10^{-8}$ mbar; (ii) the NIR detector focusing stages
remained stable within the depth of focus over the first few months;
(iii) no effects of flexure have been recognised.

First calibration observations included the derivation of zeropoints
for all seven bands, checks for vignetting and flexure, the determination
of the focus formula (see $\S$\ref{foc}), and the effect of the
K-band dither mirror.

\begin{figure*}[th]
\includegraphics[width=.8\textwidth,bb=50 50 533 764, clip]{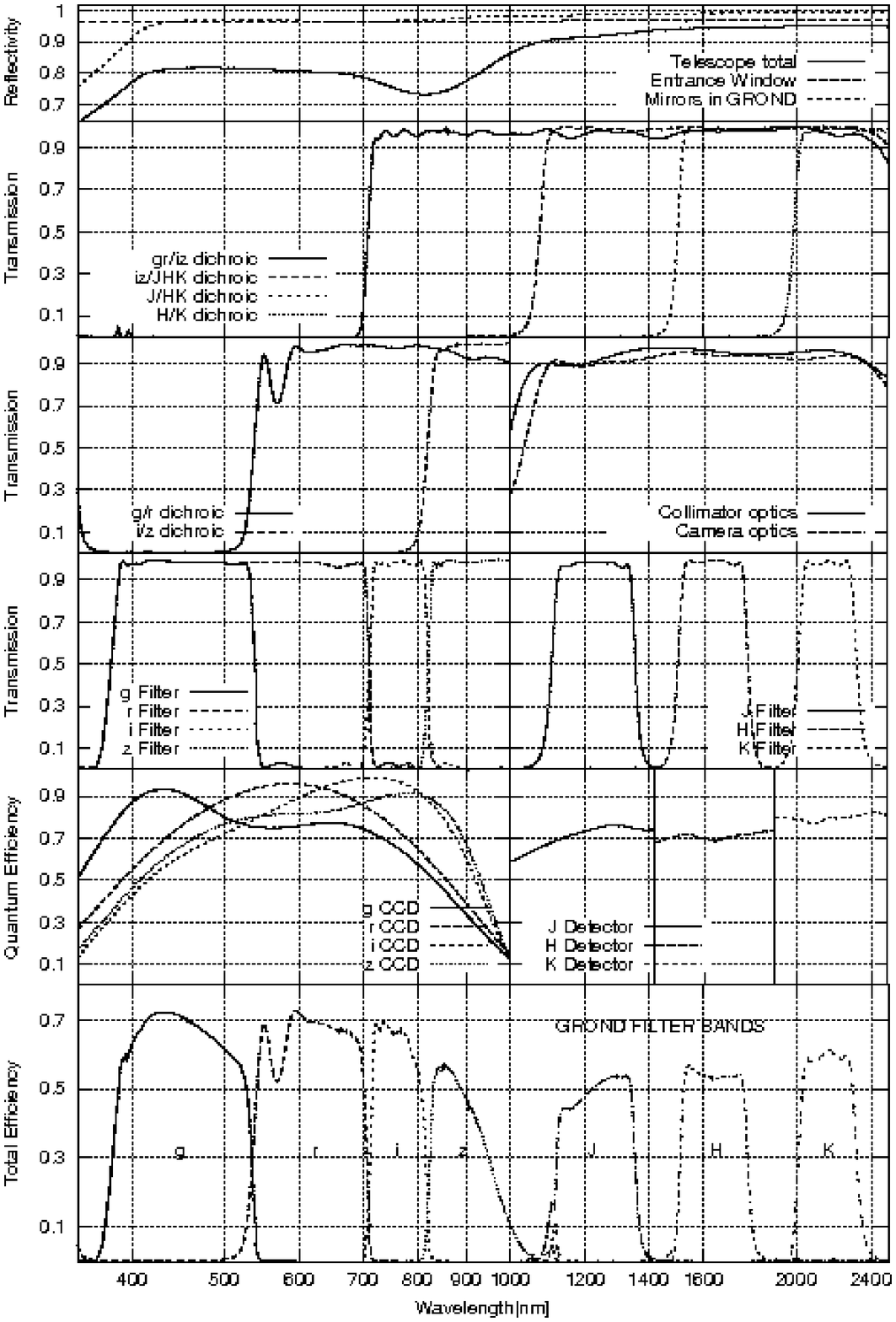}
\vspace{-0.4cm}
\caption[GROND efficiency]{Efficiency of the GROND instrument at 
the 2.2\,m telescope:
Top panel: telescope and GROND-internal mirrors.
Panels 2 and 3 from top: transmission of the dichroics, NIR lens systems,
Panel 4: Transmission of the filters.
Panel 5: Detector quantum efficiencies.
Bottom panel: Total efficiency in each of the seven GROND bands.
All losses are included, except the obstruction of M1 by M2.
Except for the telescope (M1, M2) data, all curves represent measured data
at their operating temperatures, i.e. all transmission values for the
lenses, dichroic and anti-reflection coatings refer to 80\,K.
The measured efficiency in all seven bands on the sky turns out to be 
within 10\% of the one expected from the bottom panel.}
\label{filter}
\end{figure*}

\subsection{Efficiency and limiting magnitudes \label{limmag}} 

Best possible efficiency has been a driver of many decisions
during the development of GROND, included special
sensitisation of the four CCDs for their respective wavelength
band, silver (rather than aluminum) coatings of the various mirrors, and
stringent transmission requirements of the dichroics and
anti-reflection coatings. The efficiencies of the various
elements in the optical path, including the telescope mirrors,
are shown in Fig. \ref{filter}. The total efficiency in the visual bands
is at the 70\% level (except the $z'$ band), and still above 50\% for the
three NIR bands. We note here that the combination of dichroics
and JHK filters leads to a very high efficiency in GROND's NIR part
despite the eleven lenses per channel and the comparatively low quantum
efficiency of the Hawaii detectors.

\begin{figure}[th]
\includegraphics[width=1.1\columnwidth]{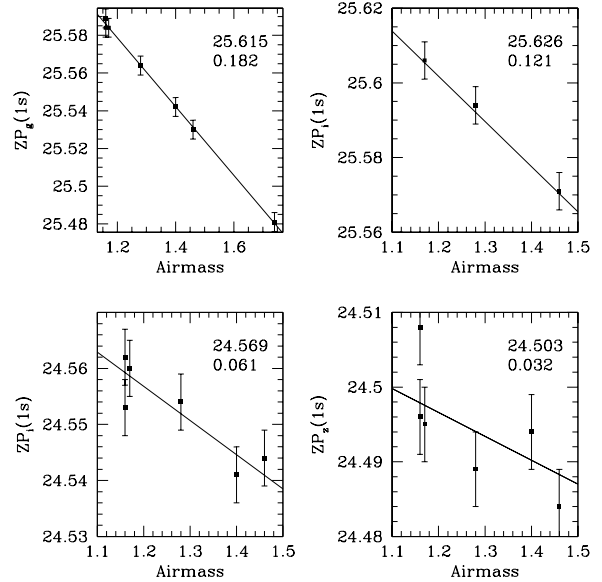}
\vspace{-0.4cm}
\caption[GROND zeropoints]{Airmass dependence of the photometric zeropoint 
in the visual bands. Magnitudes are in the Vega system and the zeropoint 
is calculated for 1\,s integration and for
ADU (i.e. not electrons). The extinction coefficients and the extrapolated
zeropoint at an airmass of 1 are shown for all four bands.
\label{zpvis}}
\end{figure}

Photometric zeropoints were determined by observing Landolt stars SA114-750,
SA114-656 in all seven channels and SA114-654 in the near-infrared at many
different airmasses \cite{landolt}. In the $r'i'z'$ bands the inherent
photometric accuracy is 0.5\% and 1\% in the $g$-band 
(magnitudes are consistent
at this level in a single OB). Long-term (spreading two photometric nights) and
airmass dependencies in all four visual bands are shown in Fig. \ref{zpvis},
demonstrating that GROND is capable to perform below 1\% accurate
photometry. The measured extinction coefficients are $0.182\pm0.003$,
$0.121\pm0.005$, $0.061\pm0.007$ and $0.044\pm0.006$ in the $g'$, $r'$, $i'$ 
and $z'$-band, respectively.

The main reason GROND can not deliver absolute photometry at this level in
the visual bands is the lack of calibration standards \emph{in the native
GROND system}. Both SA-114 stars used are primary SDSS standards 
(Smith \etal\ 2002).
The spectral types of both stars are also known (Drilling \& Landolt 1979,
Cohen \etal\ 2003). Using spectral templates from calspec/STSCI
and Pickles (1998), we estimated the magnitudes of these
two stars in the native GROND system. The combined effect of the photometric
accuracy of the primary SDSS standards and our conversion limits the
achievable overall accuracy in the $r'$, $i'$ and $z'$-band to 1\%. 
In the g-band,
the accuracy is limited to 5\%. This is due to the fact that the camera
has a significantly better response in the UV as expected. Thus, the
system becomes undefined below 3500 {\AA} and the magnitudes can not be
converted to the native system of the instrument. To demonstrate that the 
$g'$-band can perform just as accurately, we introduced an external UV filter
temporarily that cuts all UV flux. With this correction, the accuracy of
the $g'$-band is also around 1\% -- slightly worse than the other three
bands as the templates used are not sufficiently accurate around the 4000 \AA\
break. Unfortunately the optical quality of our UV filter was so bad that
it prevents scientific use. Consequently, a filter will be introduced
the camera in March 2008.

In the near-IR bands, magnitudes are only accurate to 3\% in single
(10 sec) frames. Due to the significantly lower accuracy and the
significantly lower extinction coefficients (which are water content dependent
in any case -- i.e. they are different every night) the near-IR performance
can only be determined by observing long-term trends. Currently we
accept zeropoints of 22.97, 22.22 and 21.51 in $J$, $H$ and $K$, respectively,
for 1\,sec integration time and airmass of 1. Until more data is collected,
we will use nominal extinction coefficients of 0.12 ($J$), 0.06 ($H$) and 
0.07 ($K$).

Updates of these performance values will be posted at 
www.mpe.mpg.de/\~\,jcg/GROND.

\begin{table}[th]
\caption[Limiting mag of the 7 bands]{Zeropoints and 3$\sigma$ limiting 
magnitudes for an 8 min observation block with four telescope dither
positions (corresponding to an effective 
exposure time of 4$\times$130 sec in the visual and  4$\times$12$\times$10 sec 
in the NIR channels)
for the seven GROND channels  at the 2.2\,m telescope. The error in the
zeropoint is about $\pm$0.03 mag. The limiting magnitudes in the NIR are the 
mean values; they exhibit seasonal variations by up to 
$\pm$0.5 mag. \label{ZP}}
\begin{tabular}{cccc}
\noalign{\smallskip}
\hline
GROND    & \multicolumn{2}{c}{zeropoints (mag)} & lim. magnitude \\
channel  & Vega & AB & in 8 min OB \\
\hline
g' & 25.61 & 25.62 & 24.2  \\
r' & 25.63 & 25.78 & 24.2  \\
i' & 24.57 & 24.96 & 23.7  \\
z' & 24.50 & 25.01 & 23.5  \\
J  & 22.97 & 23.88 & 21.4  \\
H  & 22.22 & 23.60 & 20.4  \\
K  & 21.51 & 23.30 & 19.0  \\
\hline
\end{tabular}
\end{table}

\subsection{Photometric system transformations}

A first set of calibration observations was performed during mid-July 2007,
though not in photometric conditions. Thus, the constants in
the following equations may change slightly 
(see www.mpe.mpg.de/\~\,jcg/GROND for updates).

First, we derive transformation equations bet\-ween the GROND (G) visual
bands and the Sloan (S) filter system:

\noindent
$ g'_{\rm G} = g'_{\rm S} + (0.000\pm0.001) 
                      + (0.105\pm0.014) \times (u'-g')_{\rm S}  
                      + (0.036\pm0.024) \times (g'-r')_{\rm S}$ \\
$ r'_{\rm G} = r'_{\rm S} + (-0.001\pm0.001)
                      + (0.056\pm0.011) \times (g'-r')_{\rm S} 
                      + (-0.089\pm0.022) \times (r'-i')_{\rm S} $ \\
$ i'_{\rm G} = i'_{\rm S} + (-0.001\pm0.001)
                      + (-0.029\pm0.007) \times (r'-i')_{\rm S} 
                      + (0.004\pm0.011) \times (i'-z')_{\rm S}$ \\
$ z'_{\rm G} = z'_{\rm S} + (-0.001\pm0.001)
                      + (0.034\pm0.003) \times (i'-z')_{\rm S}.  $

Next, the transformations between the GROND NIR bands and the 2MASS (2M)
system are as follows:

\noindent
$ J_{\rm G} =  J_{\rm 2M} + (-0.003\pm0.001)
                     + (0.005\pm0.001) \times (J-H)_{\rm 2M} $ \\
$ H_{\rm G} =  H_{\rm 2M} +  (0.002\pm0.001)
                     + (0.021\pm0.001) \times (J-H)_{\rm 2M}$ \\
$ K_{\rm G} =  K_{\rm 2M} + (-0.003\pm0.001)
                     +  (0.094\pm0.003) \times (J-K)_{\rm 2M}. $ 

\noindent
For transformation from 2MASS to other systems see Carpenter (2003).

Finally, the transformations between the visual GROND bands and the
WFI (W) system are:

\noindent
$ g'_{\rm G} = B_{\rm W} + (0.001\pm0.012)
                 + (0.132\pm0.022) \times (U-B)_{\rm W}
                 + (-0.127\pm0.053) \times (B-V)_{\rm W} $ \\
$ r'_{\rm G} = V_{\rm W} + (0.001\pm0.001)
                 + (0.039\pm0.023) \times (B-V)_{\rm W}
                 + (-0.852\pm0.024) \times (V-R)_{\rm W} $ \\
$ i'_{\rm G} = R_{\rm W} + (0.016\pm0.003)
                 + (-0.163\pm0.046) \times (V-R)_{\rm W}
                 + (-0.485\pm0.044) \times (R-I)_{\rm W} $ \\
$ z'_{\rm G} = I_{\rm W} + (0.012\pm0.002)
                 + (-0.089\pm0.005) \times (R-I)_{\rm W}.  $

\noindent
Only for the $r'$ band we reach a residual 1\% rms scatter, 
while for $i'$ and $z'$ the residual rms scatter is 2\%.
For  a transformation to the standard Landolt (Johnson $BV$ and Cousins $RI$) 
system see the dedicated WFI
Web-page\footnote{www.ls.eso.org/lasilla/sciops/2p2/E2p2M/WFI/zeropoints/}.

\section{Summary}

GROND is an imaging system capable of operating in seven colors simultaneously.
It has been designed and built at MPE Garching, and commissioned 
at the 2.2\,m MPI/ESO
telescope on La Silla, Chile. First observations show that all 
properties are according to specifications/expectations. The  first
observations of gamma-ray bursts with GROND have also been obtained
(Greiner \etal\ 2007a,b; Primak \etal\ 2007, Kr\"uhler \etal\ 2007). 
Fine tuning of the
operations strategy as well as scheduling and analysis software 
in the upcoming weeks is expected to bring GROND into a fully
operational condition, thus allowing the commencement of normal
science operations.

\bigskip

\acknowledgments
We thank K. Meisenheimer, R.-R. Rohloff and R. Wolf (all MPIA
Heidelberg) for their support in getting the telescope interfaces right,
and for FE computations of the M3 mirror rigidity.
Particular thanks to the whole La Silla Observatory staff for their
enthusiasm and effort during the assembly of all the GROND components
 to the telescope. 
We thank D.H. Hartmann (Clemson Univ.) for stimulating discussions, 
K. Garimella (formerly also Clemson Univ) and D.A. Kann 
(Tautenburg Observatory) for help in implementing 
SED-fitting routines as preparatory steps for HyperZ, 
as well as A. Rossi (Tautenburg Observatory) for support in the derivation
of the photometric conversion equations.
We are greatful to the anonymous referee for the many detailed comments
which improved to readability and consistency of the paper.
We greatly acknowledge the special efforts of the following
companies to fulfil our often unusual requirements: 
Berliner Glas KGaA,
ECM Moosinning,
FEE GmbH Idar-Oberstein,
Korth Kristalle GmbH Altenholz,
Laseroptik GmbH Garbsen,
Laser Zentrum Hannover e.V.,
Pr\"azisionsoptik Gera,
Steinbach-K\"onitzer-Lopez Jena,
Tafelmaier D\"unnschichttechnik Rosenheim.
Part of the funding for GROND (both hardware as well as personnel)
was generously granted from the Leibniz-Prize (DFG grant HA 1850/28-1)
to Prof. G. Hasinger (MPE).

\bigskip

{\it Facilities:} \facility{Max Planck:2.2m}.

\bigskip





\end{document}